\documentclass[twocolumn,journal]{IEEEtran}
\usepackage{bm}
\usepackage{amsmath,amssymb,amsfonts,graphicx,amsthm,mathtools,nicefrac,mathrsfs}
\usepackage{graphicx, subfig}
\usepackage{float}
\usepackage{times}
\usepackage{color}
\usepackage[table]{xcolor}
\usepackage{booktabs}
\usepackage{multirow}
\usepackage{lscape}

\usepackage[utf8]{inputenc}
\usepackage[T1]{fontenc}
\usepackage{url}
\usepackage{ifthen}
\usepackage{cite}
\DeclareMathOperator{\E}{\mathbb{E}}

\DeclareMathOperator{\rad}{rad}
\DeclareMathOperator{\diam}{diam}

\DeclareMathOperator{\sign}{sign}

\newcommand{\ip}[2]{\left\langle#1,#2\right\rangle}

\def \CC {\mathcal{C}}

\def \NN {\mathcal{N}}

\def \TT {\mathcal{T}}

\def \OO {\mathcal{O}}
\def \QQ {\mathcal{Q}}

\def \B {\mathbb{B}}
\def \R {\mathbb{R}}
\def \S {\mathbb{S}}

\def \va {\bm{a}}
\def \vb {\bm{b}}
\def \vd {\bm{d}}
\def \ve {\bm{e}}
\def \vg {\bm{g}}
\def \vh {\bm{h}}
\def \vn {\bm{n}}

\def \vs {\bm{s}}
\def \vu {\bm{u}}
\def \vv {\bm{v}}

\def \vx {\bm{x}}
\def \vy {\bm{y}}
\def \vz {\bm{z}}

\def \vomega {\bm{\omega}}

\def \vtau {\bm{\tau}}

\def \vzero {\bm{0}}

\def \mI {\bm{I}}

\def \mP {\bm{P}}
\def \mX {\bm{X}}
\def \mU {\bm{U}}

\def \mPhi {\bm{\Phi}}

\newtheorem{theorem}{Theorem}
\newtheorem{lemma}{Lemma}

\newtheorem{corollary}{Corollary}
\newtheorem{condition}{Condition}

\newtheorem{fact}{Fact}

\theoremstyle{definition}

\theoremstyle{remark}
\newtheorem{remark}{Remark}

\makeatletter
\newtheorem*{rep@theorem}{\rep@title}
\newcommand{\newreptheorem}[2]{%
	\newenvironment{rep#1}[1]{%
		\def\rep@title{#2 \ref{##1}}%
		\begin{rep@theorem}}%
		{\end{rep@theorem}}}
\makeatother

\newreptheorem{theorem}{Theorem}
\newreptheorem{lemma}{Lemma}
\newreptheorem{corollary}{Corollary}
\newreptheorem{proposition}{Proposition}
\begin{document}
	
\title{Quantized Corrupted Sensing with Random Dithering}

\author{Zhongxing~Sun, Wei~Cui, and Yulong~Liu	
	\thanks{The material in this paper was presented in part at the IEEE International Symposium on Information Theory, Los Angeles, USA, June 21–26, 2020 \cite{Sun2020Quantized}.}
	\thanks{Z.~Sun and W.~Cui are with the School of Information and Electronics, Beijing Institute of Technology, Beijing 100081, China (e-mail: zhongxingsun@bit.edu.cn; cuiwei@bit.edu.cn).}%
	\thanks{Y.~Liu is with the School of Physics, Beijing Institute of Technology, Beijing 100081, China (e-mail: yulongliu@bit.edu.cn).}
    \thanks{Corresponding author: Yulong Liu.}
}
\maketitle

\begin{abstract}
	Corrupted sensing concerns the problem of recovering a high-dimensional structured signal from a collection of measurements that are contaminated by unknown structured corruption and unstructured noise. In the case of linear measurements, the recovery performance of different convex programming procedures (e.g., generalized Lasso and its variants) is well established in the literature. However, in practical applications of digital signal processing, the quantization process is inevitable, which often leads to non-linear measurements. This paper is devoted to studying corrupted sensing under quantized measurements. Specifically, we demonstrate that, with the aid of uniform dithering, both constrained and unconstrained Lassos are able to recover signal and corruption from the quantized samples when the measurement matrix is sub-Gaussian. Our theoretical results reveal the role of quantization resolution in the recovery performance of Lassos. Numerical experiments are provided to confirm our theoretical results.
\end{abstract}


\section{Introduction}
Corrupted sensing is concerned with the problem of reconstructing a high-dimensional structured signal from a relatively small number of corrupted measurements \footnote{In this paper, we assume that $\mPhi$ is a sub-Gaussian sensing matrix with isotropic rows, the factor $\sqrt{m}$ in \eqref{corrupted sensing} makes the columns of $\mPhi$ and $\sqrt{m}\mI_m$ have the same scale, which helps our theoretical results to be more interpretable.}
\begin{align}\label{corrupted sensing}
\vy=\mPhi\vx^{\star}+\sqrt{m}\vv^{\star}+\vn,
\end{align}
where $\mPhi\in\R^{m\times n}$ is the measurement matrix, $\vx^{\star}\in\R^n$ and $\vv^{\star}\in\R^m$ denote the unknown structured signal and corruption respectively, and $\vn\in\R^m$ is some potential additive measurement noise. The goal is to recover $\vx^{\star}$ and $\vv^{\star}$ from given knowledge of $\vy$ and $\mPhi$. When $\vv^\star$ might contain some useful information, this model \eqref{corrupted sensing} can be interpreted as the signal separation (or demixing) problem. In particular, in the absence of corruption ($\vv^\star=\vzero$), this model \eqref{corrupted sensing} reduces to the standard compressed sensing problem.

This problem has found abundant applications in signal processing and machine learning, such as source separation \cite{elad2005simultaneous}, face recognition \cite{Wright2009Robust}, subspace clustering \cite{Elham2009Sparse}, sensor network analysis \cite{Haupt2008Compressed}, latent variable modeling \cite{Chandrasekaran2008Rank}, principle component analysis \cite{Candes2011Robust}, and so on. The performance guarantees of this problem have also been extensively investigated under different settings in the literature, important examples include sparse signal recovery from sparse corruption \cite{laska2009exact,wright2010dense,Li2013Compressed,nguyen2013exact,Nguyen2013Robust,kuppinger2012uncertainty,Studer2012Recovery,pope2013probabilistic,Studer2014Stable,zhang2018uniform}, low-rank matrix recovery from sparse corruption \cite{Chandrasekaran2008Rank, Candes2011Robust, xu2012robust, xu2013outlier,wright2013compressive,chen2013low,li2017Low}, and structured signal recovery from structured corruption \cite{Mccoy2014Sharp,Foygel2014Corrupted,Zhang2017On,Chen2017Corrupted,Jinchi2018Stable,Sun2019Recovery,Sun2021Phase}.

Although this problem is ill-posed in general, tractable recovery is achievable when both signal and corruption exhibit some low-complexity structures. Let $f(\cdot)$ and $g(\cdot)$ be some suitable norms which promote structures of signal and corruption, respectively (e.g., the $\ell_1$-norm promotes sparsity for vectors and the nuclear norm promotes low-rankness for matrices). When prior information of both signal $f(\vx^{\star})$ and corruption $g(\vv^{\star})$ is known beforehand, a natural method to recover $\vx^\star$ and $\vv^{\star}$ is the generalized constrained Lasso:
\begin{align}\label{G-Lasso}
\begin{split}
\min_{\vx, \vv} ~\|\vy-\mPhi\vx-\sqrt{m}\vv\|_2,\quad\text{s.t.~}& f(\vx)\leq f(\vx^{\star})\\
&g(\vv)\leq g(\vv^{\star}).
\end{split}
\end{align}
When there is no prior knowledge available, it is practical to use the generalized unconstrained Lasso, which solves the following penalized recovery procedure:
\begin{align}\label{f_penalized procedure}
\min_{\vx, \vv} ~\frac{1}{2}\|\vy-\mPhi\vx-\sqrt{m}\vv\|_2^2+\lambda_1\cdot f(\vx)+\lambda_2\cdot g(\vv),
\end{align}
where $\lambda_1,\lambda_2>0$ are some regularization parameters. The theoretical analyses of the above two recovery procedures under linear measurements \eqref{corrupted sensing} are well established in the literature, see e.g., \cite{Foygel2014Corrupted,Mccoy2014Sharp,Chen2017Corrupted,Zhang2017On,Jinchi2018Stable,Sun2021Phase} and references therein.

However, in the era of digital signal processing, the measurements $\vy$ are inevitably quantized into bitstreams for the
purpose of data storage and processing. So we actually obey the following non-linear observation model
\begin{equation}\label{quantized corrupted sensing}
\vy=\QQ(\mPhi\vx^{\star}+\sqrt{m}\vv^{\star}+\vn),
\end{equation}
where $\QQ(\cdot)$ stands for some quantization scheme. A fundamental problem then to ask is:

\emph{ Is it still possible to disentangle signal and corruption from the non-linear measurements? If possible, how to recover the structured signal from the quantized samples with provable performance guarantees? }

\begin{figure}[h]
	\centering
	\subfloat[Saprse signal recovery from sparse corruption]{\includegraphics[width=3in]{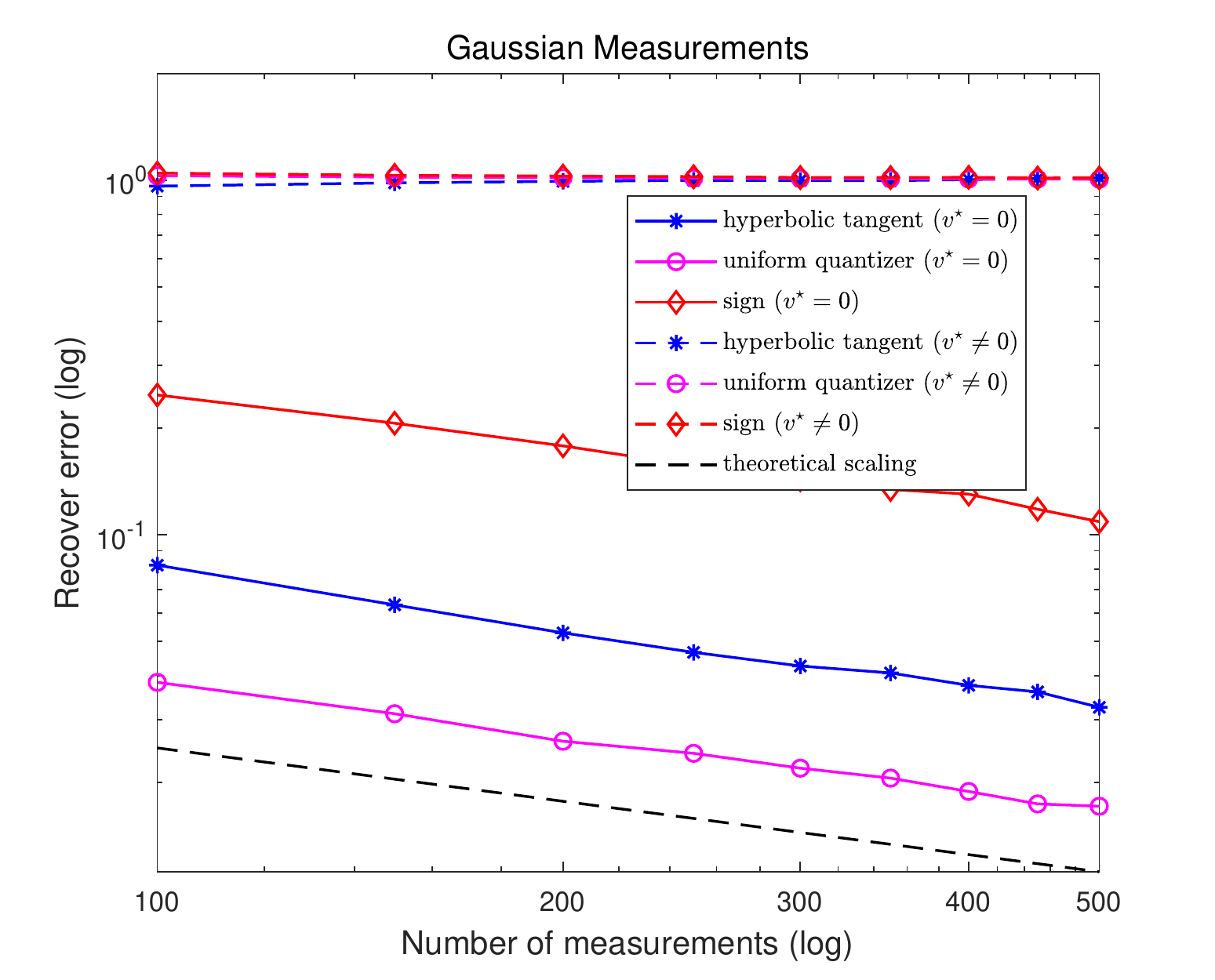}
		\label{figb}}
	\caption{Log-log error curves when use the generalized constrained Lasso \eqref{G-Lasso} to recover signal and corruption from quantized corrupted sensing measurements \eqref{quantized corrupted sensing} with different non-linear functions $\QQ(\cdot)$. The black dashed line illustrates the $\OO(\frac{1}{\sqrt m})$ error decay which is predicted in the literature (see e.g., \cite{Plan2015The}). Both signal and  corruption are  assumed to be sparse vectors. More details about this experiment can be found in Section \ref{fig_detail}.}
	\label{fig}
\end{figure}

It is now well-known that the generalized constrained Lasso developed in linear compressed sensing (without corruption $\vv$ in \eqref{G-Lasso}) also works well in the non-linear case \cite{Plan2015The}. One might naturally employ the generalized Lassos (\eqref{G-Lasso} and \eqref{f_penalized procedure}) to reconstruct signal and corruption from quantized corrupted measurements \eqref{quantized corrupted sensing}. However, direct application of \eqref{G-Lasso} and \eqref{f_penalized procedure} to measurements \eqref{quantized corrupted sensing} seems to yield unsatisfactory results. To see this, we present a numerical example to show that the generalized constrained Lasso \eqref{G-Lasso} is unable to faithfully reconstruct $\vx^\star$ from the quantized corrupted measurements \eqref{quantized corrupted sensing}. Fig. \ref{fig} displays the empirical results of recovering sparse signals from non-linear measurements \eqref{quantized corrupted sensing} under Gaussian measurements. In the corruption-free cases ($\vv^{\star}=\vzero$), the recovery errors observe a desired decay as the measurements increase, which is consistent with the theoretical prediction in \cite{Plan2015The}. Nevertheless, when the corruption appears ($\vv^{\star}\neq\vzero$), the recovery errors are relatively large and keep almost unchanged as the measurements increase, which indicates that the presence of unknown corruption $\vv^{\star}$ makes the recovery problem more challenging.

To overcome the above difficulty, we consider a specific but more tractable quantization scheme, namely dithered quantization. Dithered quantization is a technique in which a random dithering signal is added to the input before quantization. This approach is commonly used in practice because suitably chosen dithering signal can result in favorable statistical properties of the quantization error and hence more pleasing reproduction see, e.g., \cite{Schuchman1965Dither,Gray1993Dithered,Xu2018Quantized,Dirksen2018Non,Christos2018The}. In this quantization scheme, the observation process becomes
\begin{equation}\label{dither measurement model}
\vy=\QQ_U(\bar{\vy}+\vtau)=\QQ_U\left(\mPhi\vx^{\star}+\sqrt{m}\vv^{\star}+\vn+\vtau\right),
\end{equation}
where $\bar{\vy}=\mPhi\vx^{\star}+\sqrt{m}\vv^{\star}+\vn$, $\QQ_U(x):=\Delta(\lfloor\frac{x}{\Delta}\rfloor+\frac{1}{2})$ is the uniform scalar quantizer with resolution $\Delta> 0$, and $\vtau_i\sim \text{Unif}(-\frac{\Delta}{2},\frac{\Delta}{2}]$ is the uniform dithering signal. The objective is  to disentangle signal and corruption from $\mPhi$ and the quantized samples $\{\vy_i\}_{i=1}^{m}$.

\subsection{Model Assumptions and Contributions}
In this paper, we demonstrate that, by adding uniformly distributed dithering before quantization, one is able to eliminate the influence of corruption $\vv^{\star}$ and to recover original structured signal $\vx^{\star}$ from the quantized corrupted sub-Gaussian measurements \eqref{dither measurement model} via the generalized Lassos (\eqref{G-Lasso} and \eqref{f_penalized procedure}). To present our results more precisely, we require the following model assumptions:
\begin{itemize}
	\item Sub-Gaussian measurements: the rows $\mPhi_i^T$ of measurement matrix $\mPhi$ are independent, centered, isotropic sub-Gaussian vectors with $\|\mPhi_i\|_{\psi_2}\leq K$;
	\item Bounded noise: the entries of unstructured noise $\vn$ are bounded with $|\vn_i|\leq \epsilon$.	
\end{itemize}
Under the above model assumptions, the contribution of this paper is twofold:
\begin{itemize}
	\item First, we show that the constrained Lasso \eqref{G-Lasso} can disentangle signal and corruption from the dithered quantized measurements \eqref{dither measurement model} and the quantization plays a similar role to independent unstructured noise. Specifically, our analysis shows that if the number of measurements
$$m \geq c \cdot K^4[\omega^2(\TT_f(\vx^\star)\cap\S^{n-1})+\omega^2(\TT_g(\vv^\star)\cap\S^{m-1})],$$
then, with high probability, the solution to the constrained Lasso \eqref{G-Lasso} satisfies
	\begin{align*}
	&\sqrt{\|\hat{\vx} - \vx^{\star}\|_2^2+\|\hat{\vv} - \vv^{\star}\|_2^2} \leq \\
	&CK(\Delta+\epsilon)\cdot\frac{\omega(\TT_f(\vx^\star)\cap\S^{n-1})+\omega(\TT_g(\vv^\star)\cap\S^{m-1})}{\sqrt{m}},
	\end{align*}
	where $\TT_f(\vx^\star)$ (or $\TT_g(\vv^\star)$) is the tangent cone induced by $f$ (or $g$) at the true signal $\vx^\star$ (or corruption $\vv^\star$), $\omega\left(\TT_f(\vx^\star)\cap\S^{n-1}\right)$
(or $\omega\left(\TT_g(\vv^\star)\cap\S^{m-1}\right)$) is the spherical Gaussian width of this cone, which will be defined in Section \ref{Preliminaries}.
	
   \item Second, we analyze the unconstrained Lasso \eqref{f_penalized procedure} which requires neither prior information of signal $f(\vx^\star)$ nor that of corruption $g(\vv^\star)$ and seems more practical in applications. Our theoretical results indicate that, for some appropriate $\lambda_1$ and $\lambda_2$, if the number of measurements
       \begin{align*}
         m & \geq c \cdot K^4\big[\eta^2\left(\kappa\lambda_1\cdot\partial f(\vx^\star)\right)+\eta^2\left(\kappa\lambda_2\cdot\partial g(\vv^\star)\right) \\
           & ~~~~~~~~~~~~~+(\kappa\lambda_1\alpha_f)^2+(\kappa\lambda_2\alpha_g)^2\big],
       \end{align*}
       then, with high probability, the solution to the unconstrained Lasso \eqref{f_penalized procedure} satisfies
	\begin{align*}
	&\sqrt{\|\hat{\vx} - \vx^{\star}\|_2^2+\|\hat{\vv} - \vv^{\star}\|_2^2} \leq C\cdot\frac{\lambda_1\alpha_f+\lambda_2\alpha_g}{m}.
	\end{align*}
	Here, $\partial f(\vx^\star)$ (or $\partial g(\vv^\star)$) is the subdifferential of $f$ (or $g$) at the true signal $\vx^\star$ (or corruption $\vv^\star$), $\eta^2\left(\kappa\lambda_1\cdot\partial f(\vx^\star)\right)$ (or $\eta^2\left(\kappa\lambda_2\cdot\partial g(\vv^\star)\right)$) denotes the Gaussian squared distance to a scaled subdifferential, also defined in Section \ref{Preliminaries}, $\alpha_f$ and $\alpha_g$ are the compatibility constants with respect to $f$ and $g$ respectively, and $\kappa$ is an absolute constant. Moreover, our theoretical results also illustrate how to select regularization parameters in the unconstrained recovery procedure and shed some lights on the relationship between two approaches.
\end{itemize}

\subsection{Related Works}\label{related work}
 To the best of our knowledge, this non-linear corrupted sensing model \eqref{dither measurement model} seems novel in the literature. However, as a special case in which $\vv^{\star}=\vzero$, the non-linear compressed sensing problem have been intensively studied during the past decade. These works might be roughly classified into three categories: one-bit compressed sensing (CS), multi-bit CS, and general non-linear sensing model.

\emph{One-bit CS} problem was first introduced by Boufounos and Baraniuk \cite{Boufounos20081} in studying the recovery of sparse signals from single-bit measurements in the noiseless case. Since the norm information is absorbed in the sign function, it is standard to assume that $\|\vx\|_2=1$ in the one-bit CS model. In an original work \cite{jacques2013robust}, Jacques \emph{et al.} demonstrated that $\OO(s\log n)$ one-bit measurements are sufficient to recover an $s$-sparse
vector via
\begin{align}
\min_{\vx\in\R^n}\|\vx\|_0~~~\textrm{s.t.}~\vy=\textrm{sign}(\mPhi\vx),~\|\vx\|_2=1.
\end{align}
Note that the above sparsity constraint and norm constraint are non-convex, this program might be computationally intractable. Plan and Vershynin \cite{plan2013one} addressed this problem by proposing a tractable algorithm (linear programming) to reconstruct the signal from noiseless one-bit measurements. They have showed that $\OO(s\log^2(n/s))$ Gaussian measurements are sufficient to accurately recover any $s$-sparse signal. This result was also the first uniform  recovery result for one-bit CS problem. In a subsequent paper \cite{Plan2013Robust}, Plan and Vershynin considered noisy one-bit measurements and obtained both non-uniform and uniform stable recovery results by solving the following convex program
\begin{align}
\max_{\vx\in\R^n}\ip{\vy}{\mPhi\vx}~~~\textrm{s.t.}~\vx\in\TT.
\end{align}
The analyses of above works only apply to i.i.d. Gaussian measurement. As for the non-Gaussian measurements, Ai \emph{et al.} \cite{ai2014one} showed that the non-uniform recovery results in \cite{Plan2013Robust} can be generalized to sub-Gaussian case by imposing an additional assumption on the signal that it is not extremely sparse. On the other hand, the results mentioned so far all assumed that the signal has unit norm. In order to estimate the norm of signal, dithered quantization $\vy=\sign(\mPhi\vx^\star+\vtau)$ has been exploited in the context of one-bit compressed sensing. In \cite{knudson2016one}, Knudson \emph{et al.} showed that one can recover the signal norm from dithered one-bit measurements with Gaussian matrix $\mPhi$ and randomly chosen or deterministic dithering $\vtau$. In \cite{Baraniukr2014Exponential}, Baraniuk \emph{et al.} considered Gaussian measurements and Gaussian dithering and demonstrated a exponential decay of reconstruction error by choosing adaptive dithering $\vtau_i$ in a linear program. In the setting of sub-Gaussian or even heavy-tailed measurements, Dirksen and Mendelson \cite{Dirksen2018Non, dirksen2018robust} showed that, by adding uniformly distributed dithering signal before quantization, it is possible to accurately reconstruct high-dimensional structured signals from a small number of noisy one-bit measurements.

\emph{Multi-bit CS} concerns the problem of recovering high-dimensional signals from multi-bit quantized measurements. The most studied multi-bit CS model involves the uniform scalar quantizer $\QQ_U(x)=\Delta(\lfloor\frac{x}{\Delta}\rfloor+\frac{1}{2})$. Traditional compressed sensing theory solves this problem by treating the quantization error as additive noise, thus the reconstruction error hits a certain floor due to the resolution $\Delta$ of the quantizer. However, one may wish to be able to further reduce the reconstruction error by taking more measurements. In a series of works by Jacques and his collaborators \cite{jacques2015quantized, jacques2017small, jacques2017time, Xu2018Quantized}, this has been proven practical by introducing dithering signal before quantization. For example, Xu and Jacques \cite{Xu2018Quantized} considered dithered memoryless scalar quantization $\vy=\QQ_U(\mPhi\vx^\star+\vtau)$ and used a reconstruction method called projected back projection (PBP) for recovery, i.e.
\begin{align}
\hat{\vx}=\mP_\TT\left(\frac{1}{m}\mPhi^T\vy\right).
\end{align}
With benefit from uniform dithering, they established both non-uniform and uniform recovery results for any measurement matrix satisfying RIP (Restricted Isometry Property).
In a more related work \cite{Christos2018The}, Thrampoulidis and Rawat considered the uniform dithered quantization measurements. Their results showed that, under sub-Gaussian assumption, the solution to constrained Lasso \eqref{G-Lasso} satisfies (with high probability)
\begin{equation}\label{result of dMSQ}
\|\hat{\vx}-\vx^{\star}\|_2\leq C\Delta\cdot\frac{\omega\left(\TT_f(\vx^\star)\cap\S^{n-1}\right)}{\sqrt m}
\end{equation}
provided that $m\geq c_1\cdot\omega^2\left(\TT_f(\vx^\star)\cap\S^{n-1}\right)+c_2$. The analysis in \cite{Christos2018The} also illustrated that the constrained Lasso applies to one-bit dithered measurements with only a logarithmic rate loss. 

\emph{The general non-linear sensing model} only assumes that $\QQ$ is an unknown non-linear map. In this scenario, measurements can be approached with the semi-parametric single index model $\vy_i = f_i(\ip{\mPhi_i}{\vx^{\star}})$ \cite{ichimura1993semiparametric, horowitz1996direct}. In a seminal paper \cite{Plan2015The}, Plan and Vershynin considered the general non-linear model and showed that, under Gaussian measurements, the non-linear quantization process can still be treated as linear measurements, and the solution to constrained Lasso \eqref{G-Lasso} satisfies (with high probability)
\begin{equation}
\|\hat{\vx}-\mu_{f}\frac{\vx^{\star}}{\|\vx^{\star}\|_2}\|_2\leq C\cdot\frac{\omega(\TT_f(\vx^\star)\cap\S^{n-1})\sigma_{f}+\eta_{f}}{\sqrt m}
\end{equation}
provided that $m\geq c\cdot\omega^2(\TT_f(\vx^\star)\cap\S^{n-1})$. The notations $\mu_{f},\sigma_{f}$, and $\eta_{f}$ are some parameters used to characterize the non-linear function $f$.
Motivated by \cite{Plan2015The}, Thrampoulidis \emph{et  al.} \cite{Thrampoulidis2015The} demonstrated that the performance of the unconstrained Lasso with non-linear measurements is asymptotically the same as that with linear measurements. Later, Plan \emph{et al.} \cite{Plan2017High} proved that, for Gaussian $\mPhi$, approximately linear $f$, and bounded, star-shaped set $\TT$, one can estimate the direction of signal $\vx^\star$ from the PBP method. A recent work \cite{Sun2019Recovery} has also shown that the recovery from general non-linear measurements is robust to structured corruption which might be regarded as the saturation or overload error.
The above results are quite general, however, they can only estimate the direction of $\vx^{\star}$, and the analyses are built on the argument that the signal to be quantized ($\mPhi\vx^\star$) obeys zero mean Gaussian distribution. Thus they cannot trivially apply to the quantized corrupted measurements since $\vv^{\star}$ is not mean-zero and its distribution remains unknown, which might provide a potential explanation for the different numerical phenomena in Fig.\ref{fig}.

\subsection{Organization}
The remainder of the paper is organized as follows. In Section \ref{Preliminaries}, we review some preliminaries which are highly relevant to our theoretical analysis. Section \ref{result of DMSQ} is devoted to presenting our main results and related discussions. We carry out a series of numerical experiments to verify our theoretical results in Section \ref{simulation}. We conclude the paper in Section \ref{Conclusion}. All proofs are included in the Appendixes.

\section{Preliminaries}\label{Preliminaries}
In this section, we introduce some notations and facts that will be used in our analysis. Throughout the paper, $\S^{n-1}$ and $\B_2^n$  denote the unit sphere and unit ball in $\R^n$ under the $\ell_2$ norm, respectively. The notations $C, C', c_1, c_2, \textrm{etc.},$ stand for absolute constants which may differ from line to line.

\subsection{Convex Geometry}
\subsubsection{Dual norm and compatibility constant}
Let $\B_f^n:=\{\vu\in\R^n:f(\vu)\leq 1\}$ denote the unit ball in $\R^n$ under the norm $f$. The \emph{dual norm} of $f$ is defined as:
$$
f^*(\vx)=\sup_{\vu\in\B_f^n}\ip{\vu}{\vx}.
$$
The \emph{compatibility constant} between $f$ and the $\ell_2$ norm is defined as:
$$
\alpha_f:=\sup_{\vu\neq \vzero}{f(\vu)}/{\|\vu\|_2}.
$$
\subsubsection{Subdifferential}
The \emph{subdifferential} of a convex function $f$ at $\vx$ is the collection of vectors
\begin{equation*}\label{Definitionofsubdiff}
 \partial f(\vx) = \{\vu\in\R^n: f(\vx+\vd)-f(\vx)\ge\ip{\vu}{\vd}~\text{for all~} \vd\in\R^n\}.
\end{equation*}
For any $t\geq0$, we denote the scaled subdifferential as $t\cdot\partial f(\vx) = \{t\cdot\vu:\vu\in\partial f(\vx)\}$.

\subsubsection{Tangent cone}
The \emph{tangent cone} of a convex function $f$ at $\vx$ is the set of descent directions of $f$ at $\vx$
\begin{equation*}\label{DefinitionofTangentCone}
\TT_f(\vx) = \{\vu\in\R^n: ~f(\vx+t\cdot\vu)\leq f(\vx) ~ \textrm{for some}~ t > 0~\}.
\end{equation*}
%
\subsection{Geometric Measures}
The \emph{Gaussian width} and the \emph{Gaussian complexity} of a set $\TT \subset \R^{n}$ are, respectively, defined as
\begin{equation*}\label{Definition_Gaussian_width}
\omega(\TT) := \E \sup_{\vx \in \TT} \langle \vg, \vx \rangle, ~~ \textrm{where} ~~\vg\sim\NN(0,\mI_n),
\end{equation*}
and
\begin{equation*}\label{Definition_Gaussian_complexity}
\gamma(\TT) := \E \sup_{\vx \in \TT} |\langle \vg, \vx \rangle|, ~~ \textrm{where} ~~\vg\sim\NN(0,\mI_n).
\end{equation*}
These two geometric quantities are closely related\cite{Liaw2017A}:
\begin{equation}\label{Relation}
\left(\omega(\TT)+\|\vy\|_2\right)/3 \leq \gamma(\TT) \leq 2(\omega(\TT)+\|\vy\|_2)~~ \forall~\vy \in \TT.
\end{equation}
Another frequently used geometric quantity \emph{Gaussian squared distance} $\eta^2(\TT)$ of a subset $\TT\subset\R^n$ is defined as
\begin{equation*}\label{Definition_Gaussian_distance}
\eta^2(\TT) := \E \inf_{\vx \in \TT} \|\vg-\vx\|_2^2, ~~ \textrm{where} ~~\vg\sim\NN(0,\mI_n).
\end{equation*}

\subsection{High-Dimensional Probability}
A random variable $X$ is called a \emph{sub-Gaussian random variable} if the \emph{sub-Gaussian norm}
\begin{equation*}\label{Sub-Gaussian_Definition}
\|X\|_{\psi_2} = \inf\{t > 0: \E \exp(X^2/t^2) \leq 2\}
\end{equation*}
is finite. A random vector $\vx$ in $\R^n$ is \emph{sub-Gaussian random vector} if all of its one-dimensional marginals are sub-Gaussian random variables. The \emph{sub-Gaussian norm} of $\vx$ is defined as
\begin{equation*}
\|\vx\|_{\psi_2}:=\sup_{\vy\in\S^{n-1}}\big\| \ip{\vx}{\vy} \big\|_{\psi_2}.
\end{equation*}

A random vector $\vx$ in $\R^n$ is \emph{isotropic} if $\E(\vx\vx^T) = \mI_n$.

\subsection{Some Useful Facts}

In the analysis of our results, we will require some fundamental facts. The first one is an extended matrix deviation inequality, which allow us to establish a tight lower bound for the restricted singular value of our extended sensing matrix $[\mPhi, \sqrt{m}\mI_m]$.

\begin{fact}[Extended Matrix Deviation Inequality] \cite[Theorem 1]{Jinchi2018Stable}
	\label{Ext MDI}
	Let $\mPhi$ be an $m \times n$ matrix whose rows $\mPhi_i^T$ are independent centered isotropic sub-Gaussian vectors with $K = \max_i \|\mPhi_i\|_{\psi_2}$, and $\TT$ be a bounded subset of $\R^n\times\R^m$. Then
	for any $t\geq 0$, the event
	\begin{align*}
	\sup_{ (\va,\vb)\in \TT }&\left| \|\mPhi\va+\sqrt{m}\vb\|_2 - \sqrt{m}\cdot\sqrt{\|\va\|_2^2 + \|\vb\|_2^2} \right|\\
	&\leq CK^2[ \gamma(\TT) + t\cdot \rad(\TT) ]
	\end{align*}
	holds with probability at least $1-\exp(-t^2)$, where $\rad(\TT) := \sup_{\vx \in \TT}\|\vx\|_2$ denotes the radius of $\TT$.
\end{fact}
The second one is the Talagrand's Majorizing Measure Theorem which provides a convenient way to dominate a sub-Gaussian process.
\begin{fact}[Talagrand's Majorizing Measure Theorem] \cite[Theorem 2.1.1]{talagrand2006generic} or \cite[Theorem 4.1]{Liaw2017A}
	\label{Talagrand Them}
	Let $( X_{\vu} )_{\vu\in \TT}$ be a random process indexed by points in a bounded set $\TT \subset \R^{n}$. Assume that the process has sub-Gaussian increments, that is, there exists $M \geq 0$ such that
	\begin{equation*}
	\| X_{\vu} - X_{\vv} \|_{\psi_2} \leq M \|\vu-\vv\|_2 ~~~~ \text{for every} ~~ \vu,\vv \in \TT.
	\end{equation*}
	Then
	for any $t\geq 0$, the event
	\begin{equation*}\label{High_Probability_Bound}
	\sup_{\vu, \vv \in \TT} \big|X_{\vu} - X_{\vv}\big| \leq CM \left[ \omega(\TT) + t \cdot \diam(\TT) \right]
	\end{equation*}
	holds with probability at least $1- \exp(-t^2)$, where $\diam(\TT) := \sup_{\vx,\vy\in \TT}\|\vx-\vy\|_2$ denotes the diameter of $\TT$.
\end{fact}

\section{Main Results}\label{result of DMSQ}
This section is devoted to presenting performance guarantees of both constrained and unconstrained Lassos (\eqref{G-Lasso} and \eqref{f_penalized procedure}) for recovering signal and corruption from dithered quantized measurements \eqref{dither measurement model}. In Section \ref{recovery_G-Lasso}, we establish the performance guarantee for the constrained Lasso. Section \ref{recovery_U-Lasso} presents the theoretical analysis of the unconstrained Lasso.

\subsection{Recovery via the Constrained Lasso}\label{recovery_G-Lasso}
We start with analyzing the constrained Lasso \eqref{G-Lasso}. Let $(\hat{\vx}, \hat{\vv})$ be the solution to the constrained Lasso \eqref{G-Lasso}. It is not hard to check that the error vector $(\hat{\vx}-\vx^{\star},\hat{\vv}-\vv^{\star})$ belongs to the cone
$$
\CC_1:=\{(\va,\vb)\in\R^n\times\R^m:~\va\in\TT_f(\vx^\star)~\text{and}~\vb\in\TT_g(\vv^\star)\}.
$$
Then we have the following result.
\begin{theorem}[Constrained Lasso]
	\label{them: dither}
	Consider the dithered quantized measurement model \eqref{dither measurement model} in which the independent mean-zero additive noise $\vn$ satisfies $\|\vn\|_{\infty}\leq\epsilon$. Let $\Delta>0$ be the quantization resolution and $K = \max_i \|\mPhi_i\|_{\psi_2}$.
	If the number of measurements
	\begin{align}\label{NumberofMeasurements1}
	m \geq c \cdot K^4\cdot\gamma^2(\CC_1\cap\S^{n+m-1}),
	\end{align}
	then the solution to the constrained Lasso \eqref{G-Lasso} satisfies
	\begin{align*}
	\sqrt{\|\hat{\vx} - \vx^{\star}\|_2^2+\|\hat{\vv} - \vv^{\star}\|_2^2} \leq CK(\Delta+\epsilon)\cdot\frac{\gamma(\CC_1\cap\S^{n+m-1})}{\sqrt{m}}
	\end{align*}
	with probability at least $1-2\exp\{-\gamma^2(\CC_1\cap\S^{n+m-1})\}$.
\end{theorem}

\begin{remark}[The role of quantization]
	As shown in Theorem \ref{them: dither}, the parameters $\Delta$ and $\epsilon$ play the same role in the error bound. Thus we can conclude that, under dithered quantization scheme, the effect of quantization equals to independent additive noise. In the extreme case where the resolution of quantizer $\Delta\to 0$, the measurements in \eqref{dither measurement model} approach the linear situation $\vy=\mPhi\vx^{\star}+\sqrt{m}\vv^{\star}+\vn$, and as expected, Theorem \ref{them: dither} is consistent with the corrupted sensing theory \cite[Theorem 1]{Foygel2014Corrupted}, \cite[Theorem 2]{Jinchi2018Stable}.
\end{remark}

\begin{remark}[Bound $\gamma(\CC_1\cap\S^{n+m-1})$]
	To make use of Theorem \ref{them: dither}, we need to bound the Gaussian complexity $\gamma(\CC_1\cap\S^{n+m-1})$ in terms of some familiar parameters. Recall that \cite[Lemma 1]{Jinchi2018Stable}:
	\begin{align}\label{upper bound}
	&\gamma(\CC_1\cap\S^{n+m-1}) \leq \notag\\
	&\qquad 2\left[\omega(\TT_f(\vx^\star)\cap\S^{n-1})+\omega(\TT_g(\vv^\star)\cap\S^{m-1})+1\right].
	\end{align}
	The upper bounds of the Gaussian widths on the right side have been intensively studied in the literature (see e.g., \cite{chandrasekaran2012convex,Foygel2014Corrupted}).
\end{remark}

\begin{remark}[Distribution of dithering signal]
	The dithering signal is essential for the validity of our theorem. As illustrated in the proof (see Lemma \ref{quantization error} in Appendix \ref{proof of main}), uniformly distributed dithering ensures that the quantization error is mean-zero and independent of the input signal. It is of great interest to explore other dithering distribution when the mean-zero independent property still holds. This question is well-studied in the context of dithered quantizers, see e.g. \cite{Gray1993Dithered}.
\end{remark}

\begin{remark}[Related works]
	Observe that if we denote $\bm{\Upsilon} = [\mPhi, \sqrt{m}\mI_m]$ and $\vs^{\star} = [(\vx^{\star})^T, (\vv^{\star})^T]^T$, then the corrupted sensing model \eqref{corrupted sensing} can be reformulated as the standard compressed sensing model $\vy = \bm{\Upsilon}\vs^{\star} + \vn$. Under this observation model, Xu and Jacques \cite{Xu2018Quantized} studied the effect of dithering in the uniform quantization scheme and demonstrated that the PBP method can be employed for the recovery of $\vs^\star$ provided that the measurement matrix satisfies RIP. Note that Fact \ref{Ext MDI} implies that $\frac{1}{\sqrt{m}}\bm{\Upsilon}$ satisfies the RIP with high probability, then PBP can be naturally applied to our problem settings. However, this current paper considers totally different recovery procedures (constrained and unconstrained Lassos) and establishes corresponding error bounds which depend on quantization resolution, noise level, the number of measurements, and structures of signal and corruption. Moreover, as illustrated in Section \ref{performance_comparisons}, the constrained Lasso shows a much better recovery performance than PBP.
	
	In the corruption-free and noise-free case, i.e., $\vv^{\star} =\vzero,~\epsilon=0$,  Theorem \ref{them: dither} states that the reconstruction error $\|\hat{\vx} - \vx^{\star}\|_2 \leq CK\Delta\cdot\frac{\gamma(\TT_f(\vx^\star)\cap\S^{n-1})}{\sqrt{m}}$ provided that $m \geq c \cdot K^4\cdot\gamma^2(\TT_f(\vx^\star)\cap\S^{n-1})$. Applying the relationship between Gaussian complexity and Gaussian width \eqref{Relation}, this result reduces to \eqref{result of dMSQ}. Thus Theorem \ref{them: dither} generalizes the result in \cite[Theorem 3.1]{Christos2018The} to a more challenging scenario in which the observations might be contaminated by both structured corruption and random noise.
\end{remark}

\subsubsection{Examples}
Here, we consider two typical structured signal recovery problems: sparse signal recovery from sparse corruption and low-rank matrix recovery from sparse corruption.  To apply Theorem \ref{them: dither} in these two cases, it suffices to bound the corresponding Gaussian widths of specific structures  in \eqref{upper bound}. These results are well-established in the literature and summarized in Table \ref{table_bound}.
\begin{table}[ht]
	\caption{Closed form upper bounds for $\omega^2(\TT_f(\vx^\star)\cap\S^{n-1})$. The results are due to \cite{chandrasekaran2012convex,Foygel2014Corrupted}.}
	\centering
	\label{table_bound}
	\begin{tabular}{m{3cm}|l|c}
		\hline
		Structures & $~~f(\cdot)~~$  & Closed form upper bounds \\
		\hline
		$s$-sparse $n$-dimensional vector  & $~~\|\cdot\|_1~~$ & $2s\log(\frac{n}{s})+\frac{3}{2}s$ \\
		\hline
		$\rho$-rank $d\times d$ matrix  & $~~\|\cdot\|_*~~$ & $3\rho(2d-\rho)$ \\
		\hline
	\end{tabular}
\end{table}

\paragraph{Sparse signal recovery from sparse corruption}
In this case, assume that the signal $\vx^{\star}\in\R^n$ is $s$-sparse and that the corruption $\vv^{\star}\in\R^m$ is $k$-sparse. We use the $\ell_1$-norm, namely $f(\cdot)=g(\cdot)=\|\cdot\|_1$, to promote the structures of both signal and corruption.  Thus we have
\begin{align*}
\omega^2(\TT_f(\vx^\star)\cap\S^{n-1})\leq 2s\log\left(\frac{n}{s}\right)+\frac{3}{2}s,
\end{align*}
and
\begin{align*}
\omega^2(\TT_g(\vv^\star)\cap\S^{m-1})\leq 2k\log\left(\frac{m}{k}\right)+\frac{3}{2}k.
\end{align*}
Combining the upper bound \eqref{upper bound} and Theorem \ref{them: dither} yields the following corollary.
\begin{corollary}
	Suppose that the signal $\vx^\star\in\R^n$ is an $s$-sparse vector and that the corruption $\vv^\star\in\R^m$ is a $k$-sparse vector. Under the assumptions of Theorem \ref{them: dither}, we have that, if the number of measurements
	$$
	m\geq c \cdot K^4\left(s\log\left(\frac{n}{s}\right)+k\log\left(\frac{m}{k}\right)\right),
	$$
	then, with high probability,
	\begin{align*}
	&\sqrt{\|\hat{\vx} - \vx^{\star}\|_2^2+\|\hat{\vv} - \vv^{\star}\|_2^2} \\
	&\qquad\leq CK(\Delta+\epsilon)\cdot\frac{\sqrt{s\log(n/s)}+\sqrt{k\log(m/k)}}{\sqrt m}.
	\end{align*}
\end{corollary}

\paragraph{Low-rank matrix recovery from sparse corruption}
We then consider the case in which $\vx^{\star}=\text{vec}(\mX^{\star})\in\R^n$, where $\mX^{\star}\in\R^{d\times d}$ is a square matrix with rank $\rho$ and $d^2=n$, and the corruption $\vv^{\star}$ is a $p$-sparse vector.
We use the nuclear norm and the $\ell_1$-norm to promote the structures of signal and corruption respectively, namely, $f(\cdot)=\|\cdot\|_*$ and $g(\cdot)=\|\cdot\|_1$. In this case, we have
\begin{align*}
\omega^2(\TT_f(\vx^\star)\cap\S^{n-1})\leq 3\rho(2d-\rho),
\end{align*}
and
\begin{align*}
\omega^2(\TT_g(\vv^\star)\cap\S^{m-1})\leq 2p\log\left(\frac{m}{p}\right)+\frac{3}{2}p.
\end{align*}
Then by the upper bound \eqref{upper bound} and Theorem \ref{them: dither}, we have the following corollary.%
\begin{corollary}
	Suppose that the signal $\mX^\star\in\R^{d\times d}$ is a $\rho$-rank matrix with $d^2=n$ and that the corruption $\vv^\star\in\R^m$ is a $p$-sparse vector. Under the assumptions of Theorem \ref{them: dither}, we have that, if the number of measurements
	$$
	m\geq c \cdot K^4\left(\rho d+p\log\left(\frac{m}{p}\right)\right),
	$$
	then, with high probability,
	\begin{align*}
	&\sqrt{\|\hat{\mX} - \mX^{\star}\|_F^2+\|\hat{\vv} - \vv^{\star}\|_2^2} \\
	&\qquad\leq CK(\Delta+\epsilon)\cdot\frac{\sqrt{\rho d}+\sqrt{p\log(m/p)}}{\sqrt m}.
	\end{align*}
\end{corollary}

\subsection{Recovery via Unconstrained Lasso}\label{recovery_U-Lasso}
In this subsection, we study the recovery performance of unconstrained Lasso. Due to presence of quantization error, measurement noise and random dithering, we require the regularization parameters $\lambda_1,~\lambda_2$ to satisfy the following condition.
\begin{condition}\label{condition2}
	Let $\vz=\QQ_U(\bar{\vy}+\vtau)-\bar{\vy}-\vtau$ be the quantization error. The regularization parameters satisfy:
	\begin{align*}
	\lambda_1\geq 2 f^*\big(\mPhi^T(\vz+\vn+\vtau)\big)~\textrm{and}~\lambda_2 \geq 2\sqrt{m} g^*(\vz+\vn+\vtau).
	\end{align*}
\end{condition}
This condition is a natural extension of \cite[Theorem 11.1]{hastie2015statistical}, where only the regularization parameter for signal $\lambda_1$ is considered with $f(\cdot)=\|\cdot\|_1$. 

Let $(\hat{\vx}, \hat{\vv})$ be the solution to the unconstrained Lasso \eqref{f_penalized procedure}. Similarly, we define the following convex cone in which the error vector $(\hat{\vx}-\vx^{\star},\hat{\vv}-\vv^{\star})$ lives:
\begin{align*}
	\CC_2:=&\left\{(\va,\vb)\in\R^n\times\R^m:\lambda_1\ip{\va}{\vu}+\lambda_2\ip{\vb}{\vs}\right.\\
	&\quad\leq\frac{1}{2}[\lambda_1f(\va)+\lambda_2g(\vb)]\\
	&\left.\quad\textrm{for any }\vu\in\partial f(\vx^\star)~\text{and}~\vs\in \partial g(\vv^\star)\right\}.
\end{align*}
Then we have the following result.
\begin{theorem}[Unconstrained Lasso]
	\label{them: dither_FP}
	Consider the dithered quantized measurement model \eqref{dither measurement model} in which the independent mean-zero additive noise $\vn$ satisfies $\|\vn\|_{\infty}\leq\epsilon$. Let $\Delta>0$ be the quantization resolution and $K = \max_i \|\mPhi_i\|_{\psi_2}$. Suppose that the regularization parameters $\lambda_1,\lambda_2$ satisfy Condition \ref{condition2}. If the number of measurements
	\begin{align}\label{NumberofMeasurements3}
	m \geq c \cdot K^4\cdot\gamma^2(\CC_2\cap\S^{n+m-1}),
	\end{align}
	then the solution to unconstrained Lasso \eqref{f_penalized procedure} satisfies
	\begin{align*}
	&\sqrt{\|\hat{\vx} - \vx^{\star}\|_2^2+\|\hat{\vv} - \vv^{\star}\|_2^2} \leq C\cdot\frac{\lambda_1\alpha_f+\lambda_2\alpha_g}{m}
	\end{align*}
	with probability at least $1-\exp\{-\gamma^2(\CC_2\cap\S^{n+m-1})\}$.
\end{theorem}

\begin{remark}[Identify the ranges of $\lambda_1$ and $\lambda_2$]
Since Theorem \ref{them: dither_FP} depends on Condition \ref{condition2}, it is necessary to identify the ranges of regularization parameters $\lambda_1$ and $\lambda_2$ in our model. From Step 3 of the proof of Theorem \ref{them: dither}, we know that the quantization error $\vz_i\sim\textrm{Unif}(-\Delta/2,\Delta/2]$ are i.i.d. and also independent of $\mPhi$ and $\vn$. Then it follows Lemma \ref{lemma: UI} in Appendix \ref{proof of main} that (by setting $\vomega=\vz+\vn+\vtau,~\TT=\B_f^n\times \vzero,~t=\gamma(\B_f^n)/r_{B_f}$), the event
\begin{align*}
f^*\big(\mPhi^T(\vz+\vn+\vtau)\big)&=\sup_{\vu\in \B_f^n}\ip{\mPhi \vu }{\vz+\vn+\vtau}\\
&\leq CK(\Delta+\epsilon)\sqrt{m} \gamma(\B_f^n)
\end{align*}
holds with probability at least
$1-\exp\{-\gamma^2(\B_f^n)/r_{B_f}^2\}$, where $\B_f^n=\{\vu\in\R^n:f(\vu)\leq 1\}$ and $r_{B_f}=\sup\{\|\vu\|_2:\vu\in\B_f^n\}$. Then we can choose that
\begin{align}\label{id of lbd1}
\lambda_1\geq 2CK(\Delta+\epsilon)\sqrt{m} \gamma(\B_f^n),
\end{align}
which ensure the first part of Condition \ref{condition2} holds with high probability. On the other hand, Lemma \ref{lemma: UI} also implies (by setting $\vomega=\vz+\vn+\vtau,~\TT=\vzero\times\B_g^m ,~t=\gamma(\B_g^m)/r_{B_g}$), the event
\begin{align*}
\sqrt{m}g^*(\vz+\vn+\vtau)&=\sup_{\vu\in \B_g^m}\ip{\sqrt{m} \vu }{\vz+\vn+\vtau}\\
&\leq CK(\Delta+\epsilon)\sqrt{m} \gamma(\B_g^m)
\end{align*}
holds with probability at least
$1-\exp\{-\gamma^2(\B_g^m)/r_{B_g}^2\}$, where $\B_g^m=\{\vu\in\R^m:g(\vu)\leq 1\}$ and $r_{B_g}=\sup\{\|\vu\|_2:\vu\in\B_g^m\}$. Then we can choose that
\begin{align}\label{id of lbd2}
\lambda_2\geq 2CK(\Delta+\epsilon)\sqrt{m} \gamma(\B_g^m),
\end{align}
which ensure the second part of Condition \ref{condition2} holds with high probability.
\end{remark}

\begin{remark}[On the error decay of unconstrained Lasso]
	Since the Gaussian complexities $\gamma(\B_f^n)$ and $\gamma(\B_g^m)$ are usually much smaller than the ambient dimensions of signal and corruption, inequations \eqref{id of lbd1} and \eqref{id of lbd2} imply that we should choose the parameters $\lambda_1$ and $\lambda_2$ at least of order $\OO(\sqrt{m})$. On the other hand, Theorem \ref{them: dither_FP} illustrates that we should choose $\lambda_1$ and $\lambda_2$ as small as possible in order to achieve the possibly smallest recovery error. Thus we can achieve an error decay $\OO(\frac{1}{\sqrt{m}})$ by setting $\lambda_1$ and $\lambda_2$ to their lower bounds in \eqref{id of lbd1} and \eqref{id of lbd2} respectively.
\end{remark}

\begin{remark}[Bound $\gamma(\CC_2\cap\S^{n+m-1})$]\label{bound of gamma}
	To bound the Gaussian complexity $\gamma(\CC_2\cap\S^{n+m-1})$ in terms of familiar parameters, we make use of the following result \cite[Lemma 3]{Jinchi2018Stable}\footnote{The original lemma is slightly different from our upper bound \eqref{upper bound2} , where we have replaced $\lambda_1$ and $\lambda_2$ by $\kappa\lambda_1$ and $\kappa\lambda_2$ respectively due to the scale-invariance of $\CC_2$.}:
		\begin{align}\label{upper bound2}
		&\gamma(\CC_2\cap\S^{n+m-1})\leq\notag\\
		&2\left[\sqrt{\eta^2\left({\kappa\lambda_1}\cdot\partial f(\vx^\star)\right)+\eta^2\left({\kappa\lambda_2}\cdot\partial g(\vv^\star)\right)} \right. \notag \\
		&\qquad\qquad\qquad\qquad\qquad\left. +\frac{{\kappa\lambda_1\alpha_f+\kappa\lambda_2\alpha_g}}{2} +1\right],
		\end{align}
		where $\kappa > 0$ is a positive constant. 
	The upper bounds of the Gaussian squared distances to a scaled subdifferential on the right side have also been studied in the literature (see e.g., \cite[Appendix H]{oymak13The}).
\end{remark}

\begin{remark}[The influence of quantization and noise]
	The effects of $\Delta$ and $\epsilon$ on the reconstruction error and sample complexity are connected by the regularization parameters $\lambda_1, \lambda_2$. It follows from \eqref{id of lbd1} and \eqref{id of lbd2} that large quantization step $\Delta$ and noise level $\epsilon$ will lead to large $\lambda_1, \lambda_2$, and hence result in large recovery error in Theorem \ref{them: dither_FP}. On the other hand, observe that the definition of $\CC_2$ depends on the regularization parameters through the ratio $\lambda_2/\lambda_1$, so is the sample complexity $\gamma^2(\CC_2\cap\S^{n+m-1})$. If we set $\lambda_1$ and $\lambda_2$ to their lower bounds in \eqref{id of lbd1} and \eqref{id of lbd2} respectively, then the sample complexity is independent of $\Delta$ and $\epsilon$. To summarize, Theorem \ref{them: dither_FP} states that, similar to the constrained case, large quantization step and noise level will result in large recovery error of the unconstrained Lasso, but have litter influence on the number of measurements that required for a robust recovery provided that $\lambda_1$ and $\lambda_2$ are properly selected.
\end{remark}

\begin{remark}[Related works]
   In the context of non-linear compressed sensing, Thrampoulidis \emph{et  al.} \cite{Thrampoulidis2015The} considered the unconstrained Lasso (without corruption $\vv$ in \eqref{f_penalized procedure}) for recovery. They obtained asymptotically precise reconstruction error and demonstrated that the performance of the unconstrained Lasso with non-linear measurements is asymptotically the same as that with linear measurements. To the best of our knowledge, the guarantees in Theorem \ref{them: dither_FP} provide the first non-asymptotic theoretical results for the unconstrained Lasso \eqref{f_penalized procedure} under non-linear sensing model.
\end{remark}

\subsubsection{Examples}\label{concrete ep2}
To illustrate Theorem \ref{them: dither_FP}, we also consider two typical structured signal recovery problems: sparse signal recovery from sparse corruption and low-rank matrix recovery from sparse corruption.  To apply Theorem \ref{them: dither_FP}, it suffices to bound the Gaussian squared distances in \eqref{upper bound2} and to choose parameters $\lambda_1,\lambda_2$ for specific structures of signal and corruption. Related upper bounds for the Gaussian squared distances are summarized in Table \ref{table_bound2}.
\begin{table}[ht]
	\caption{Closed form upper bounds for $\eta^2\left(\lambda\cdot\partial f(\vx^\star)\right)$. The results are due to \cite[Appendix H]{oymak13The}.}
	\centering
	\label{table_bound2}
	\begin{tabular}{m{3cm}|l|c}
		\hline
		Structures & $f(\cdot)$  & Closed form upper bounds \\
		\hline
		$s$-sparse $n$-dimensional vector  & $\|\cdot\|_1$ & $(\lambda^2+3)s$ for $\lambda\geq \sqrt{2\log\frac{n}{s}}$\\
		\hline
		$\rho$-rank $d\times d$ matrix  & $\|\cdot\|_*$ & $\lambda^2\rho+2d(\rho+1)$ for $\lambda\geq 2\sqrt{d}$ \\
		\hline
	\end{tabular}
\end{table}

\paragraph{Sparse signal recovery from sparse corruption}

Consider the case in which the signal $\vx^{\star}\in\R^n$ is $s$-sparse and the corruption $\vv^{\star}\in\R^m$ is $k$-sparse. We have the compatibility constants $\alpha_f=\sqrt{s}$ and $\alpha_g=\sqrt{k}$. Note that \cite[Exercise 7.5.9]{Vershynin2018}
\begin{align*}
\gamma(\B_f^n)=\gamma(\B_1^n)\leq c\sqrt{\log n},
\end{align*}
and
\begin{align*}
\gamma(\B_g^m)=\gamma(\B_1^m)\leq c\sqrt{\log m}.
\end{align*}
Then, \eqref{id of lbd1} and \eqref{id of lbd2} suggest that we should pick
\begin{align*}
\lambda_1\geq C_1K(\Delta+\epsilon)\sqrt{m\log n},
\end{align*}
and
\begin{align*}
\lambda_2\geq C_1K(\Delta+\epsilon)\sqrt{m\log m}.
\end{align*}

To properly use the upper bounds in Table \ref{table_bound2}, we choose the positive constant $\kappa $ in \eqref{upper bound2} as $\kappa= {2}/(C_1K(\Delta+\epsilon)\sqrt{m})$. Clearly, we have $\kappa{\lambda_1}\geq\sqrt{2\log\frac{n}{s}}$ and $\kappa{\lambda_2}\geq\sqrt{2\log\frac{m}{k}}$, and hence
\begin{align*}
\eta^2\left(\kappa\lambda_1\cdot\partial \|\vx^\star\|_1\right)\leq \left({\kappa^2}{\lambda_1^2}+3\right)s,
\end{align*}
and
\begin{align*}
\eta^2\left(\kappa\lambda_2\cdot\partial \|\vv^\star\|_1\right)\leq \left({\kappa^2}{\lambda_2^2}+3\right)k.
\end{align*}
Substituting the above upper bounds into \eqref{upper bound2}, we obtain that
$$
\gamma^2(\CC_2\cap\S^{n+m-1}) \leq C' {\kappa^2}\cdot({\lambda_1^2s+\lambda_2^2k})= C''\cdot\frac{\lambda_1^2s+\lambda_2^2k}{K^2(\Delta+\epsilon)^2m}.
$$
Thus we have following corollary.

\begin{corollary}\label{corollary_uncon_sparse_recovery}
	Suppose that the signal $\vx^\star\in\R^n$ is an $s$-sparse vector and that the corruption $\vv^\star\in\R^m$ is a $k$-sparse vector. Let the regularization parameters satisfy $\lambda_1\geq C'K(\Delta+\epsilon)\sqrt{m\log n}$ and $\lambda_2\geq C''K(\Delta+\epsilon)\sqrt{m\log m}$. Under the assumptions of Theorem \ref{them: dither_FP}, we have that, if the number of measurements
	$$
	m\geq c \cdot K\big({\lambda_1\sqrt{s}+\lambda_2\sqrt{k}}\big)/\big(\Delta+\epsilon\big),
	$$
	then, with high probability,
	\begin{align*}
	\sqrt{\|\hat{\vx} - \vx^{\star}\|_2^2+\|\hat{\vv} - \vv^{\star}\|_2^2}\leq C\cdot\frac{\lambda_1\sqrt{s}+\lambda_2\sqrt{k}}{m}.
	\end{align*}
\end{corollary}

\begin{figure*}[!h]
	\centering
	\subfloat[Sparse signal recovery from sparse corruption]{\includegraphics[width=2.3in]{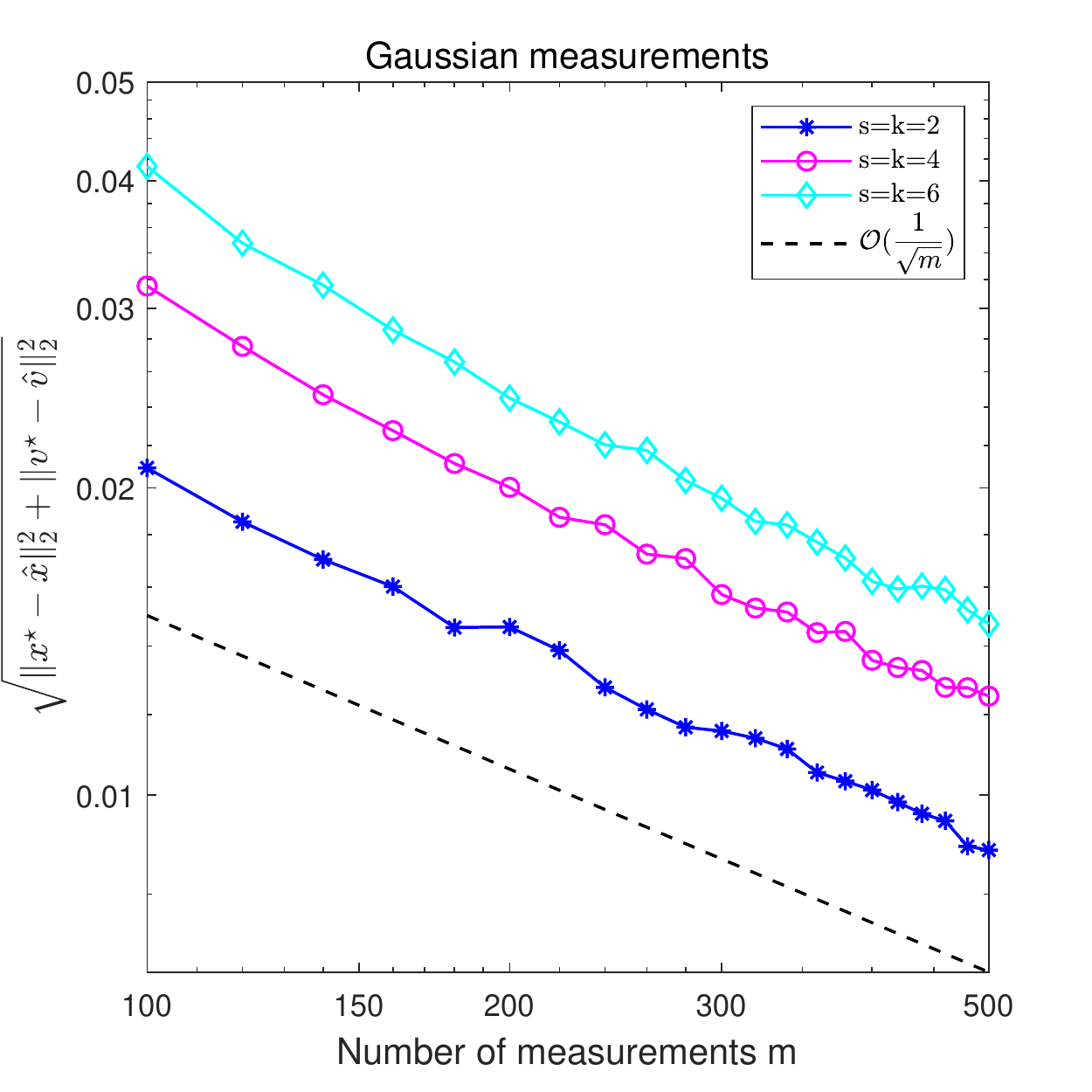}
		\label{fig1a}}
	\hfil
	\subfloat[Low-rank matrix recovery from sparse corruption]{\includegraphics[width=2.3in]{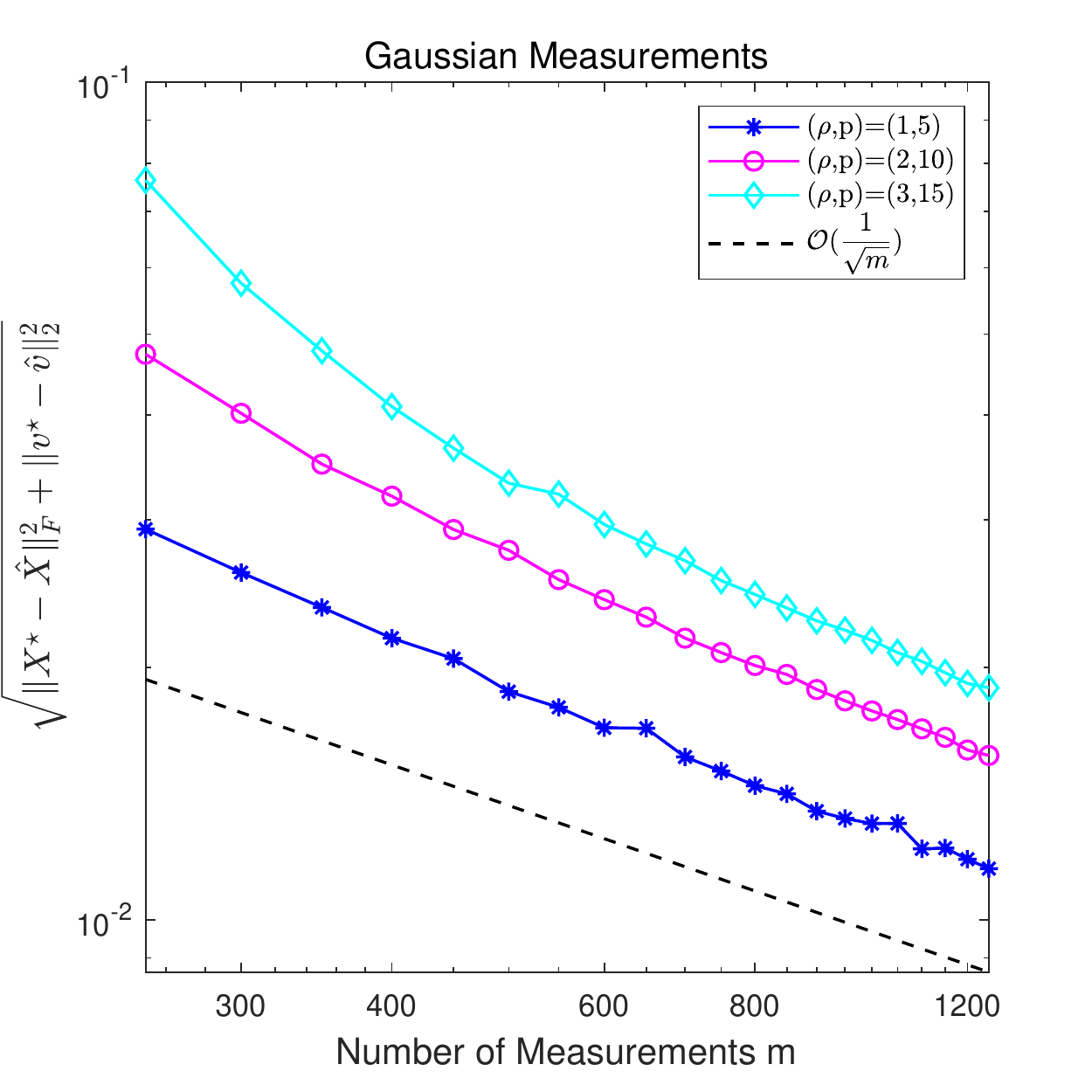}
		\label{fig1b}}
	\hfil
	\subfloat[Robustness to noise]{\includegraphics[width=2.3in]{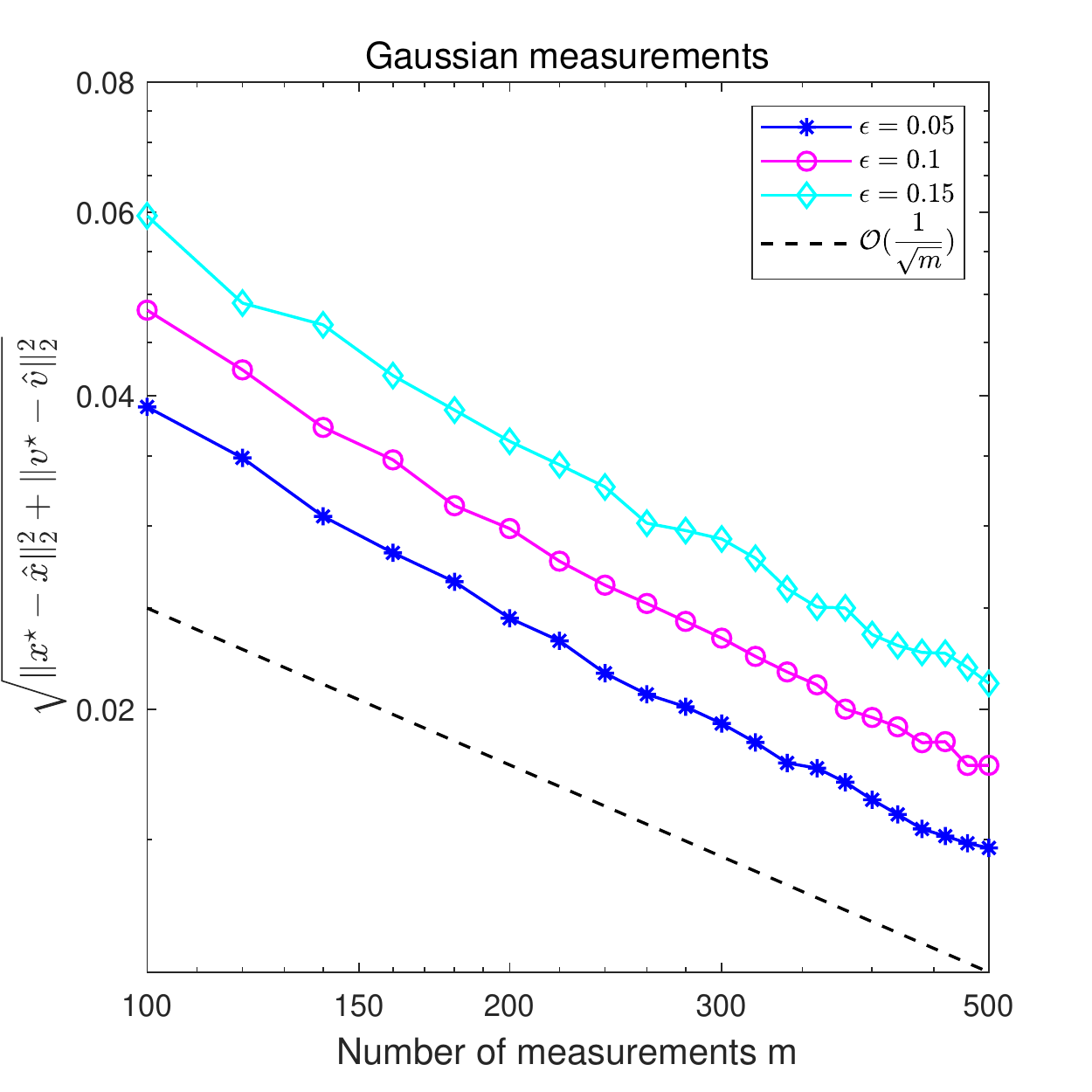}
		\label{fig1c}}
	\hfil
	\subfloat[Sparse signal recovery from sparse corruption]{\includegraphics[width=2.3in]{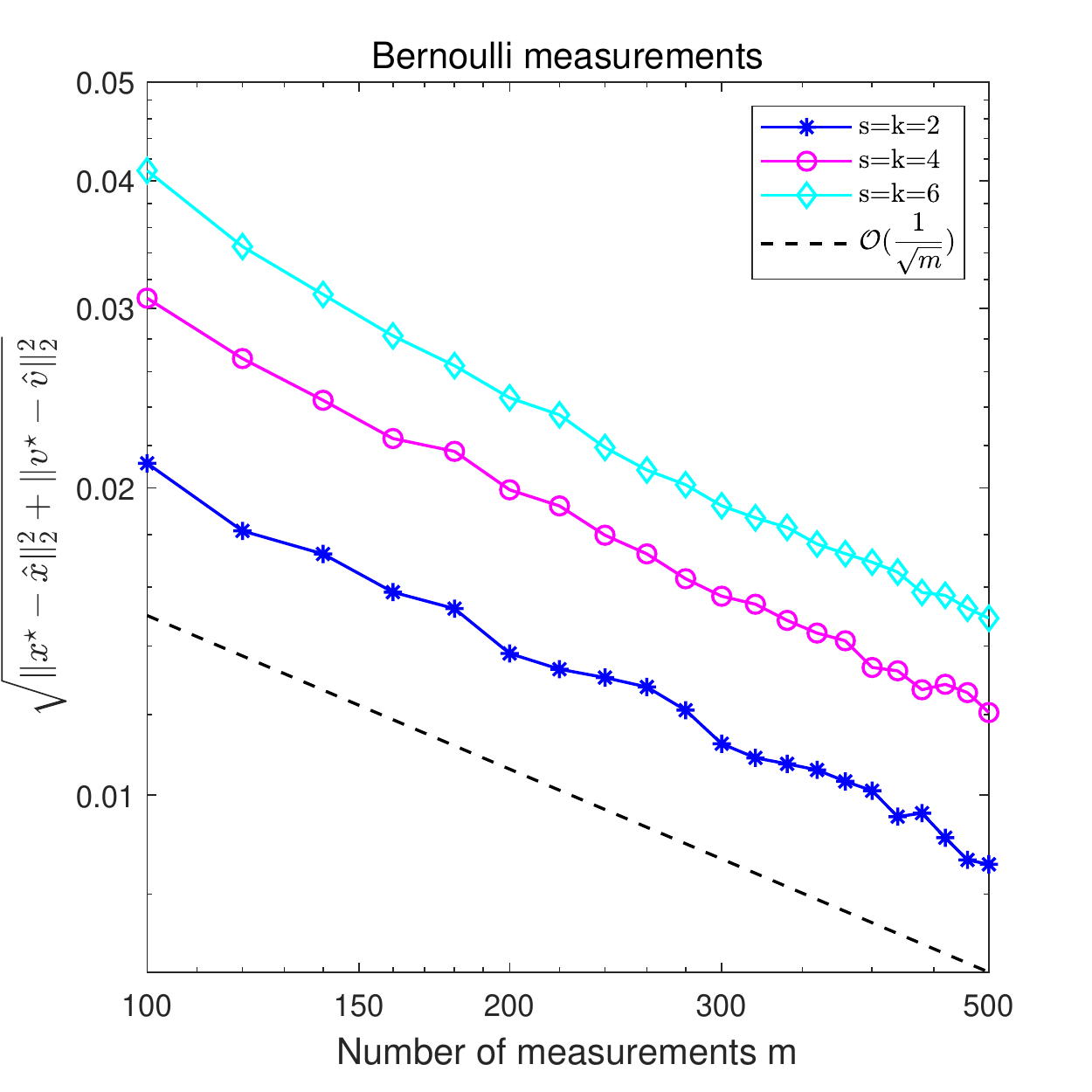}
		\label{fig1d}}
	\hfil
	\subfloat[Low-rank matrix recovery from sparse corruption]{\includegraphics[width=2.3in]{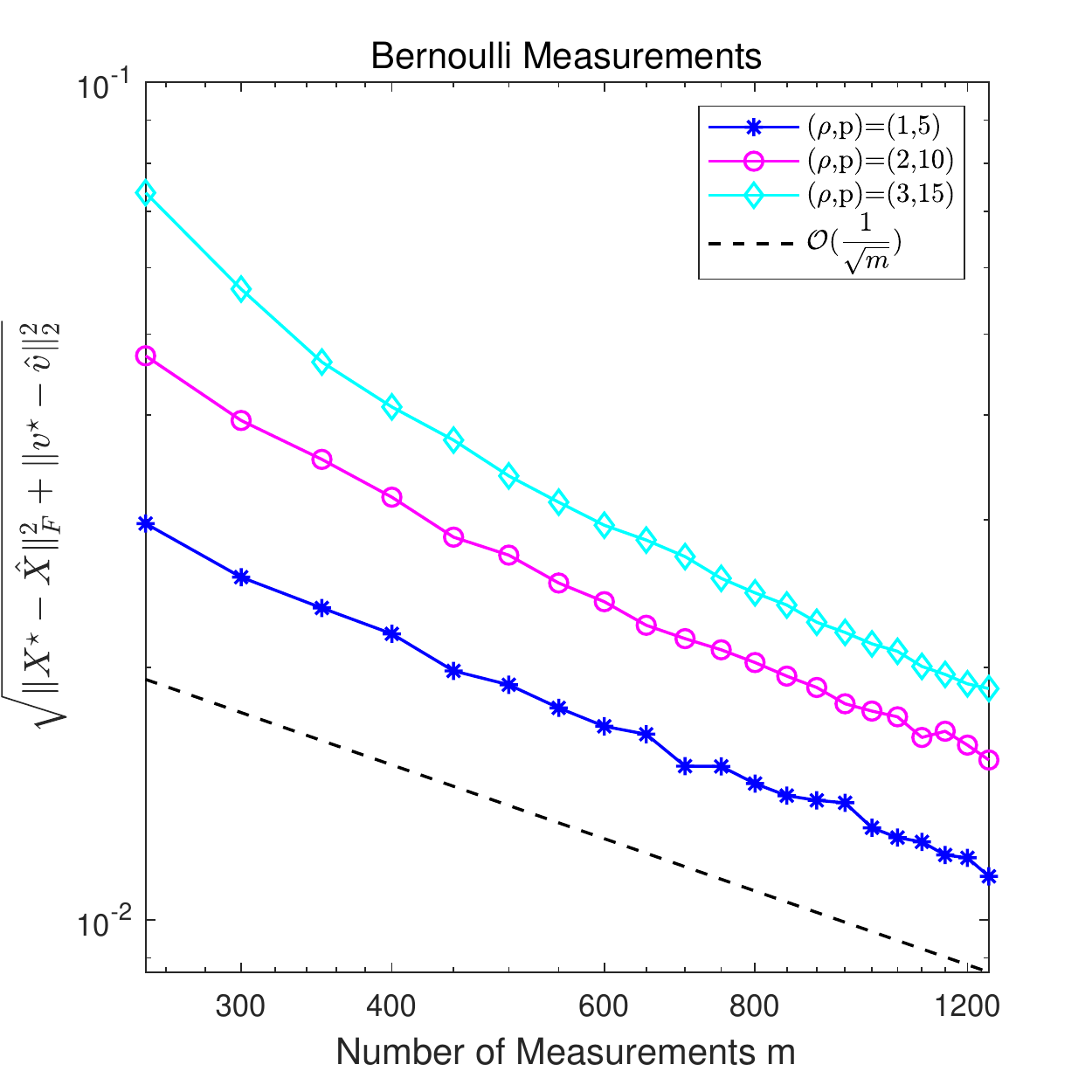}
		\label{fig1e}}
	\hfil
	\subfloat[Robustness to noise]{\includegraphics[width=2.3in]{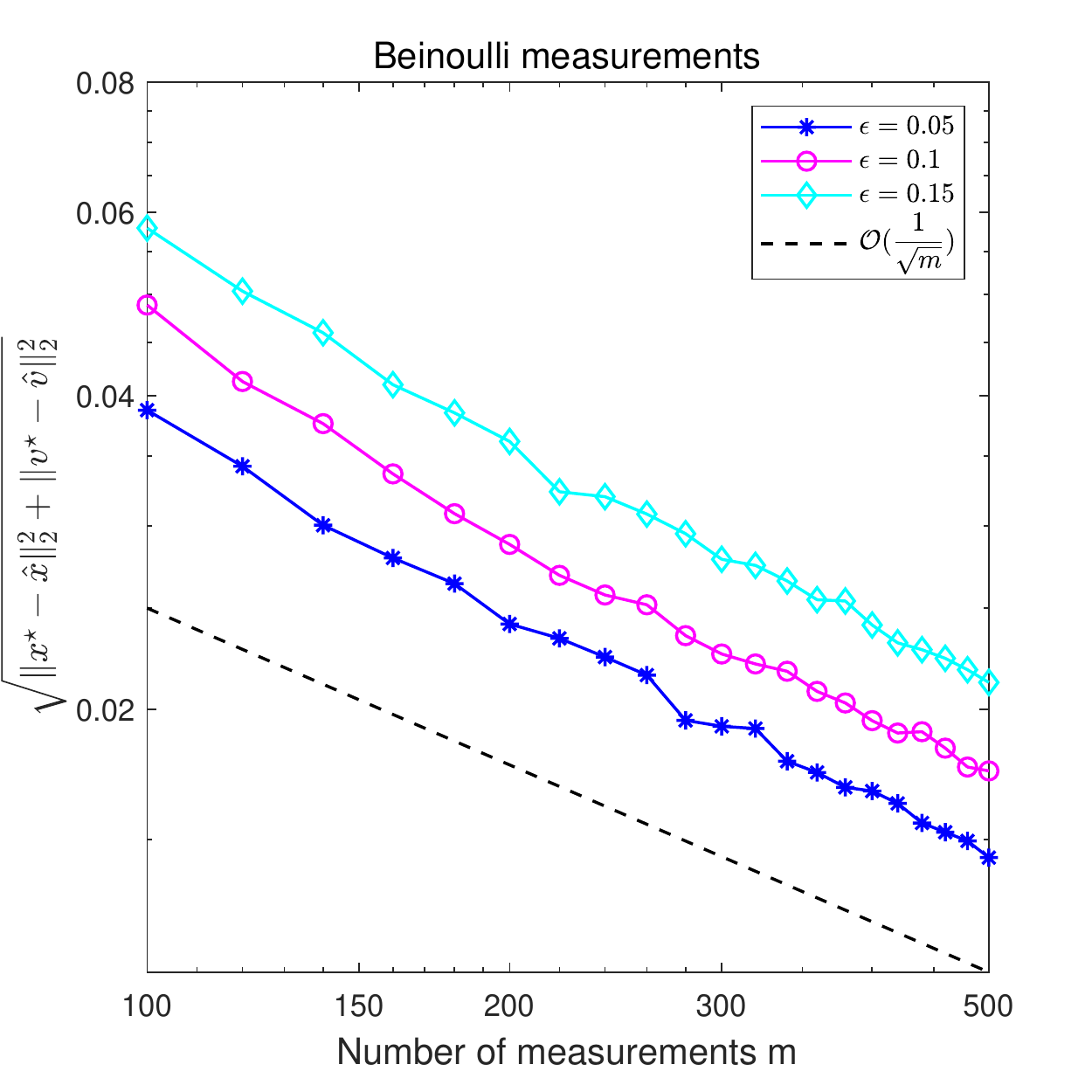}
		\label{fig1f}}
	\caption{Log-log error curves for the constrained Lasso under Gaussian or Bernoulli measurements. The black dashed line illustrates the $\OO(\frac{1}{\sqrt m})$ error decay which is predicted by Theorem \ref{them: dither}.}
	\label{fig1}
\end{figure*}

\begin{figure*}[!h]
	\centering
	\subfloat[Sparse signal recovery from sparse corruption]{\includegraphics[width=2.3in]{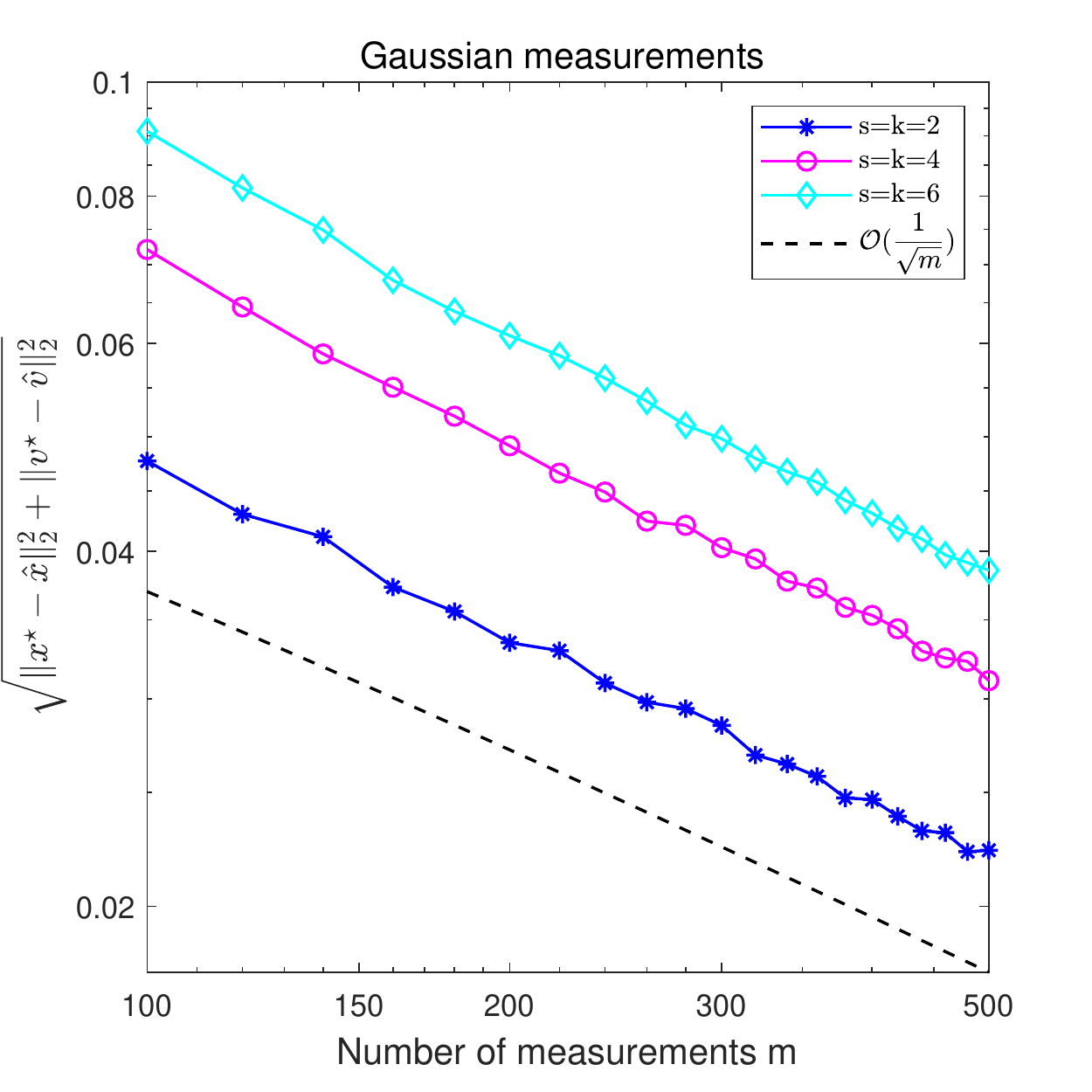}
		\label{fig2a}}
	\hfil
	\subfloat[Low-rank matrix recovery from sparse corruption]{\includegraphics[width=2.3in]{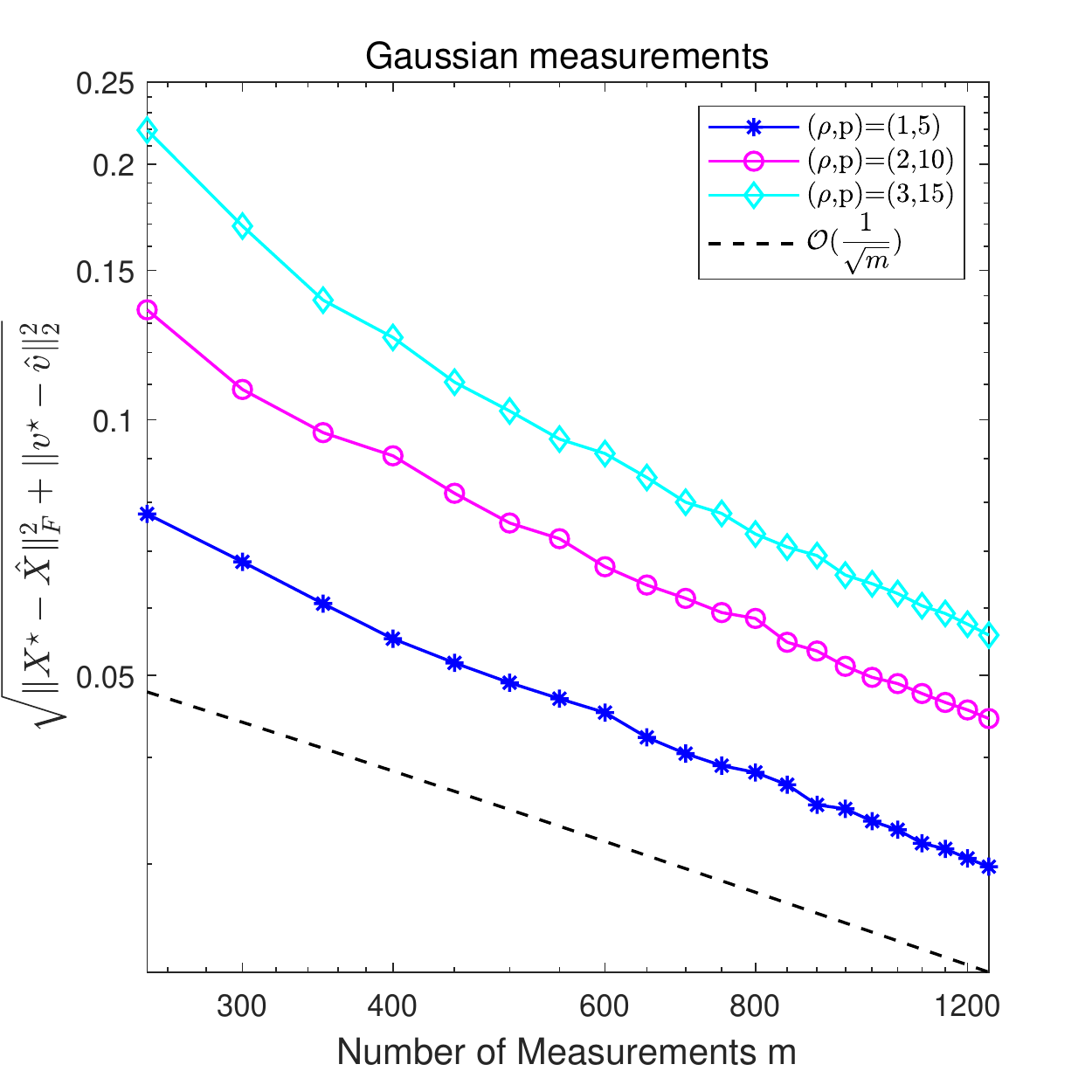}
		\label{fig2b}}
	\hfil
	\subfloat[Robustness to noise]{\includegraphics[width=2.3in]{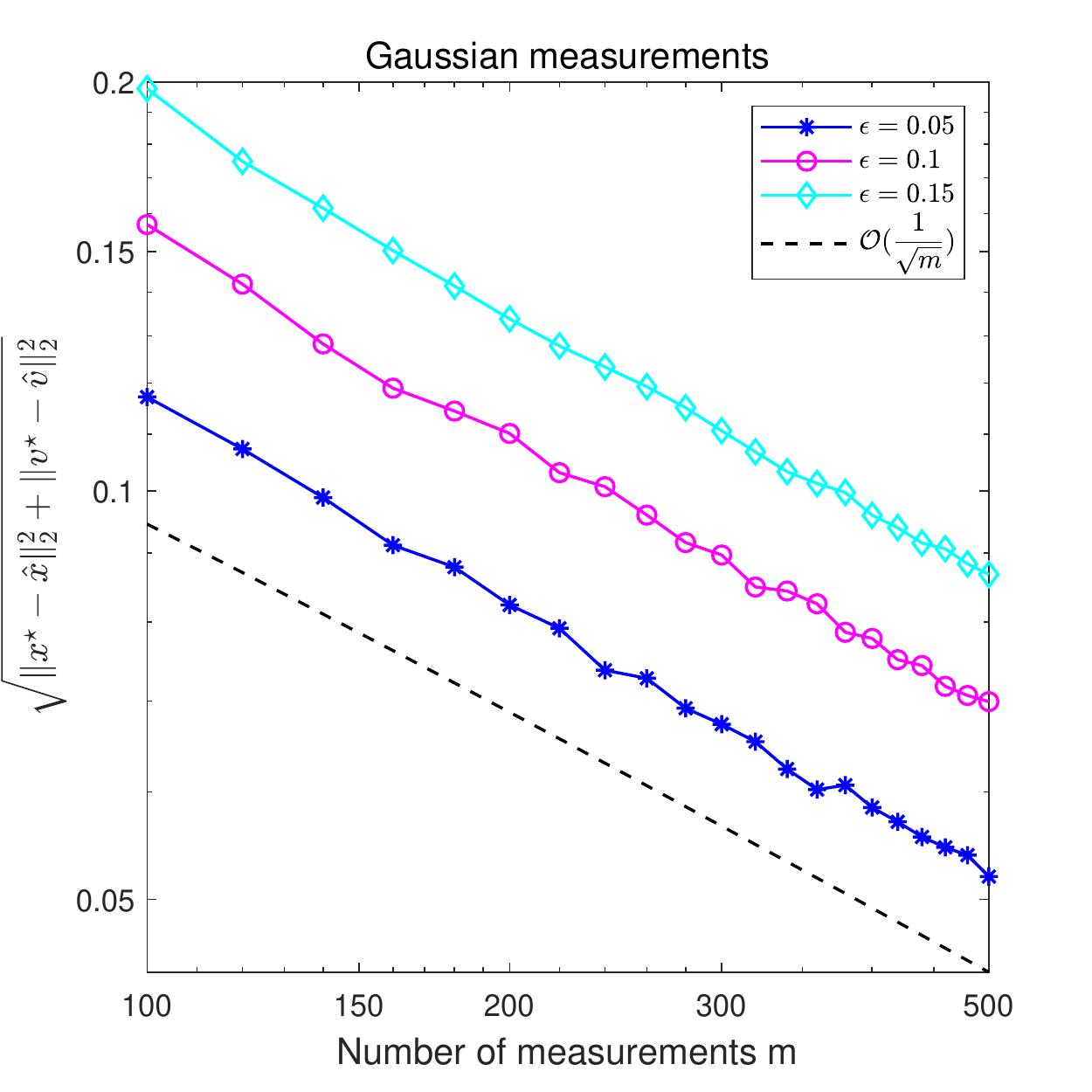}
		\label{fig2c}}
	\hfil
	\subfloat[Sparse signal recovery from sparse corruption]{\includegraphics[width=2.3in]{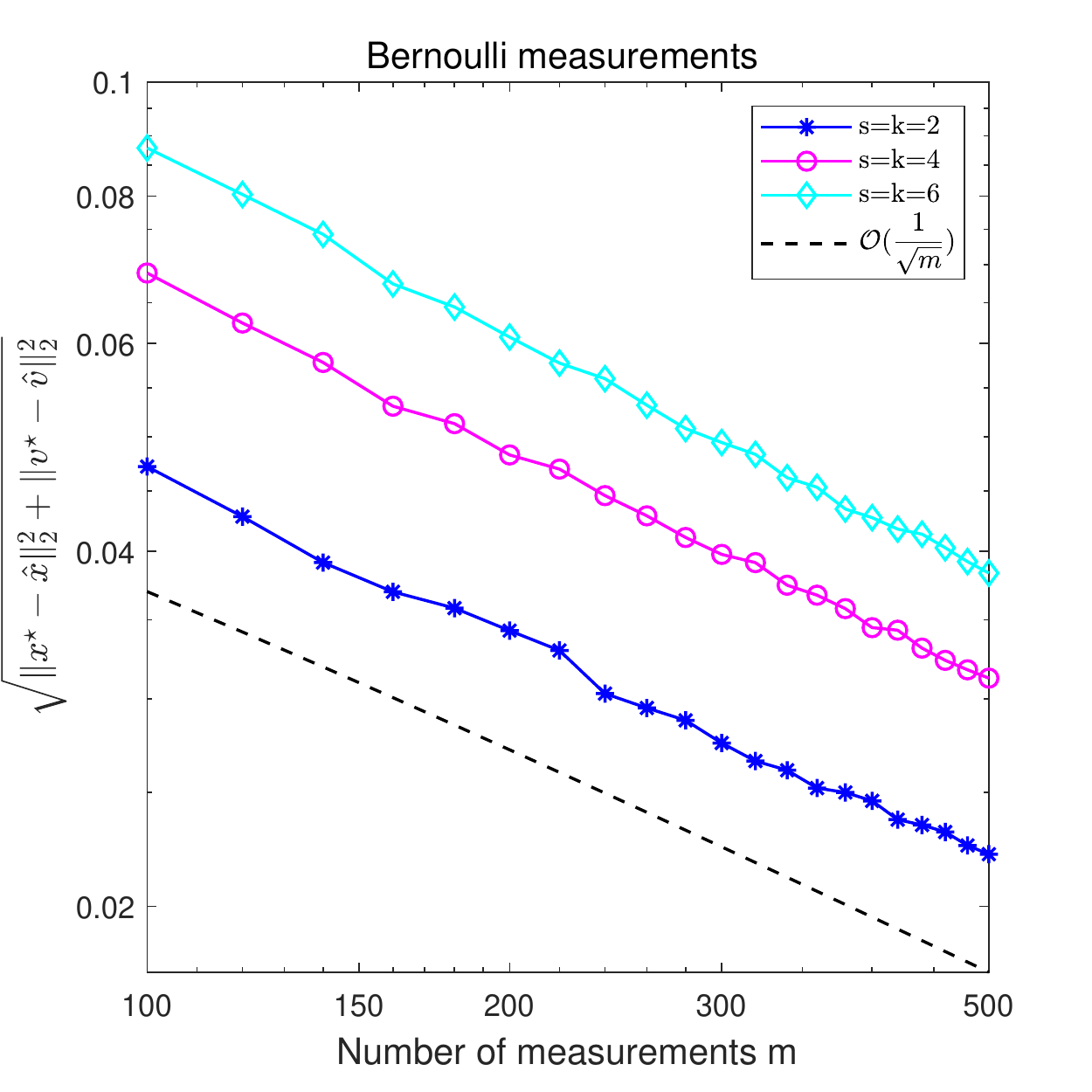}
		\label{fig2d}}
	\hfil
	\subfloat[Low-rank matrix recovery from sparse corruption]{\includegraphics[width=2.3in]{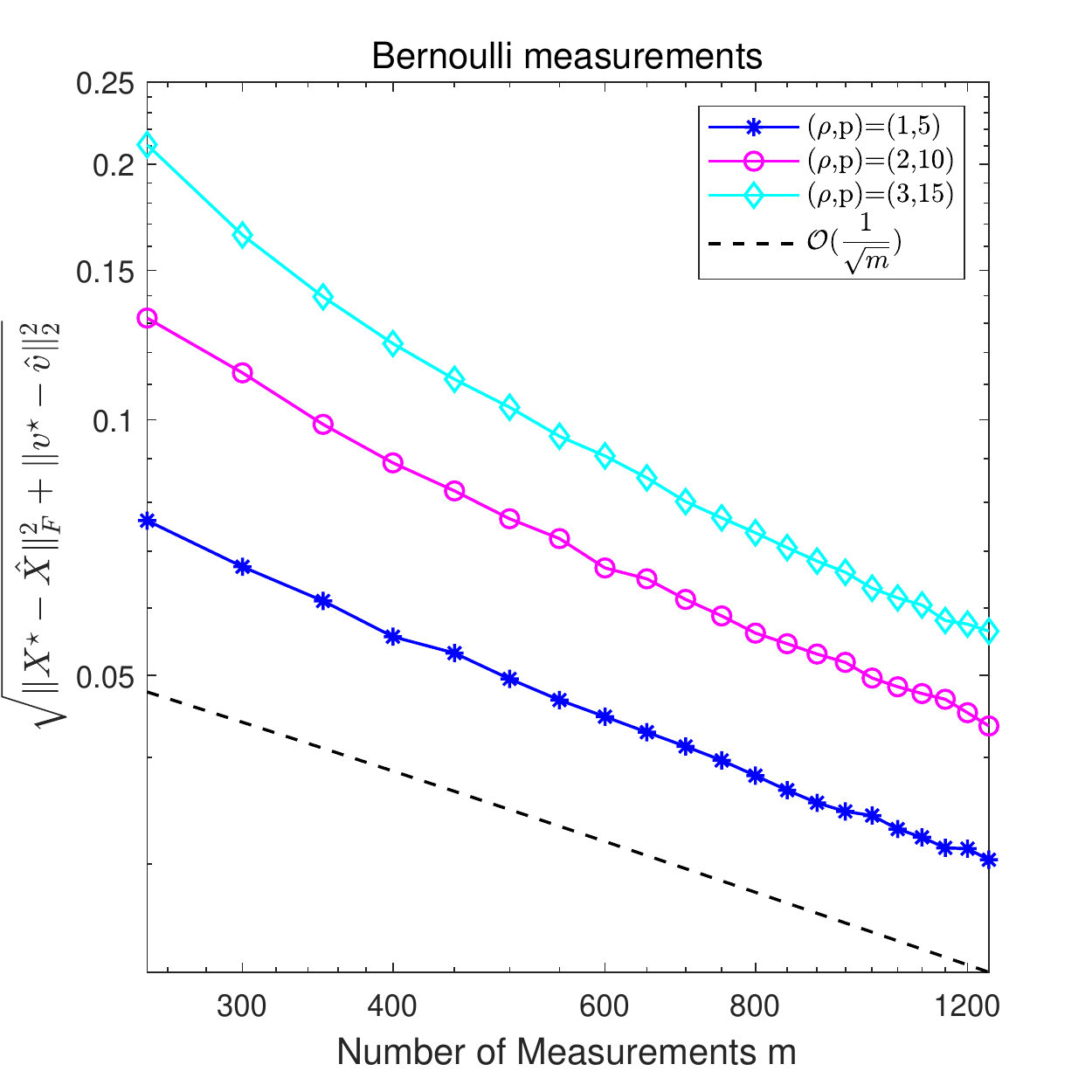}
		\label{fig2e}}
	\hfil
	\subfloat[Robustness to noise]{\includegraphics[width=2.3in]{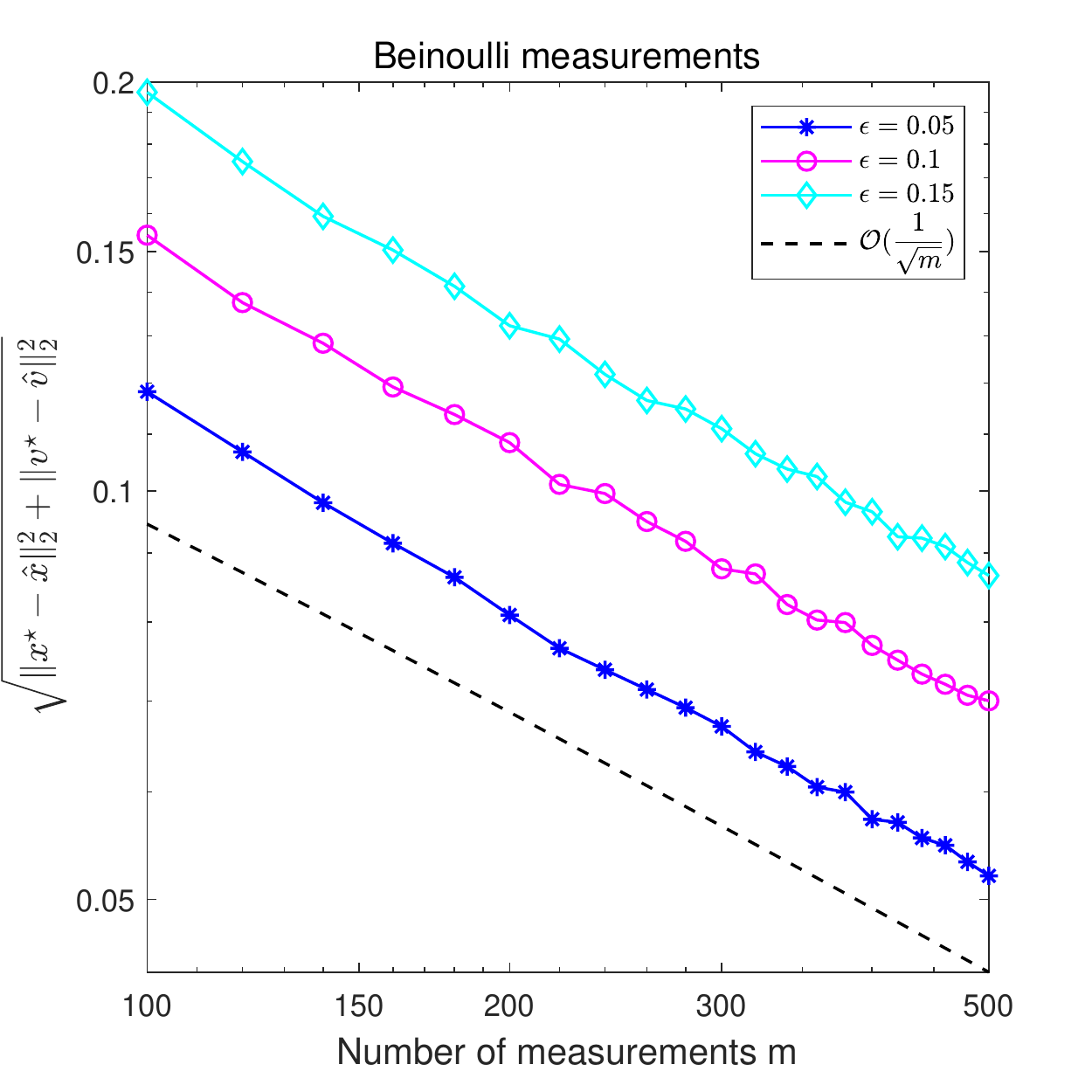}
		\label{fig2f}}
	\caption{Log-log error curves for the unconstrained Lasso under Gaussian or Bernoulli measurements. We choose $\lambda_1=(\Delta+\epsilon)\sqrt{m\log n}$, $\lambda_2=(\Delta+\epsilon)\sqrt{m\log {m}}$ for sparse signal recovery from sparse corruption, and $\lambda_1=2(\Delta+\epsilon)\sqrt{md}$, $\lambda_2=(\Delta+\epsilon)\sqrt{m\log{m}}$ for low-rank matrix recovery from sparse corruption. The black dashed line illustrates the $\OO(\frac{1}{\sqrt m})$ error decay which is predicted by Theorem \ref{them: dither_FP}.}
	\label{fig2}
\end{figure*}

\begin{figure*}[!h]
	\centering
	\subfloat[Sparse signal recovery from sparse corruption]{\includegraphics[width=2.0in]{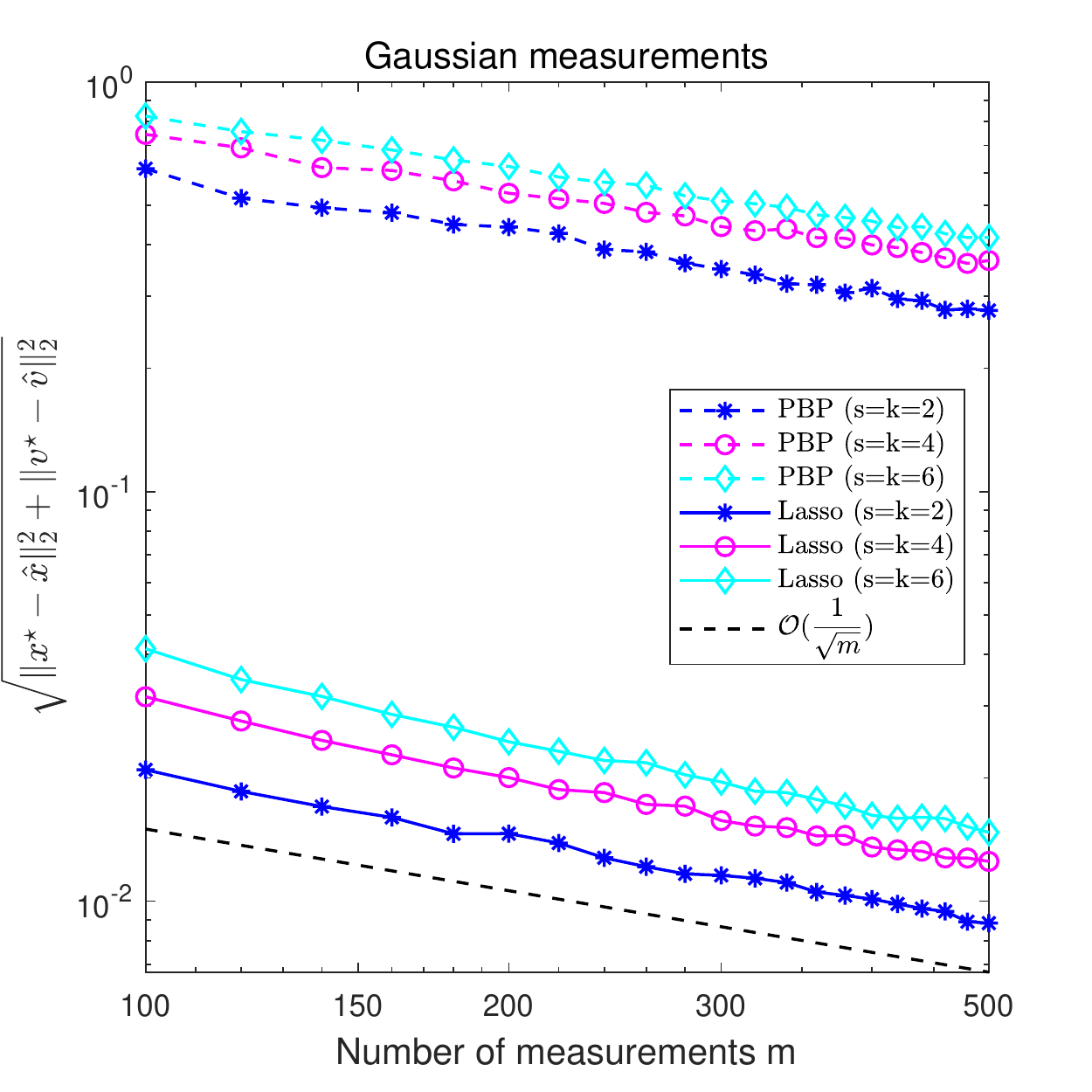}
		\label{fig3a}}
	\hfil
	\subfloat[Low-rank matrix recovery from sparse corruption]{\includegraphics[width=2.0in]{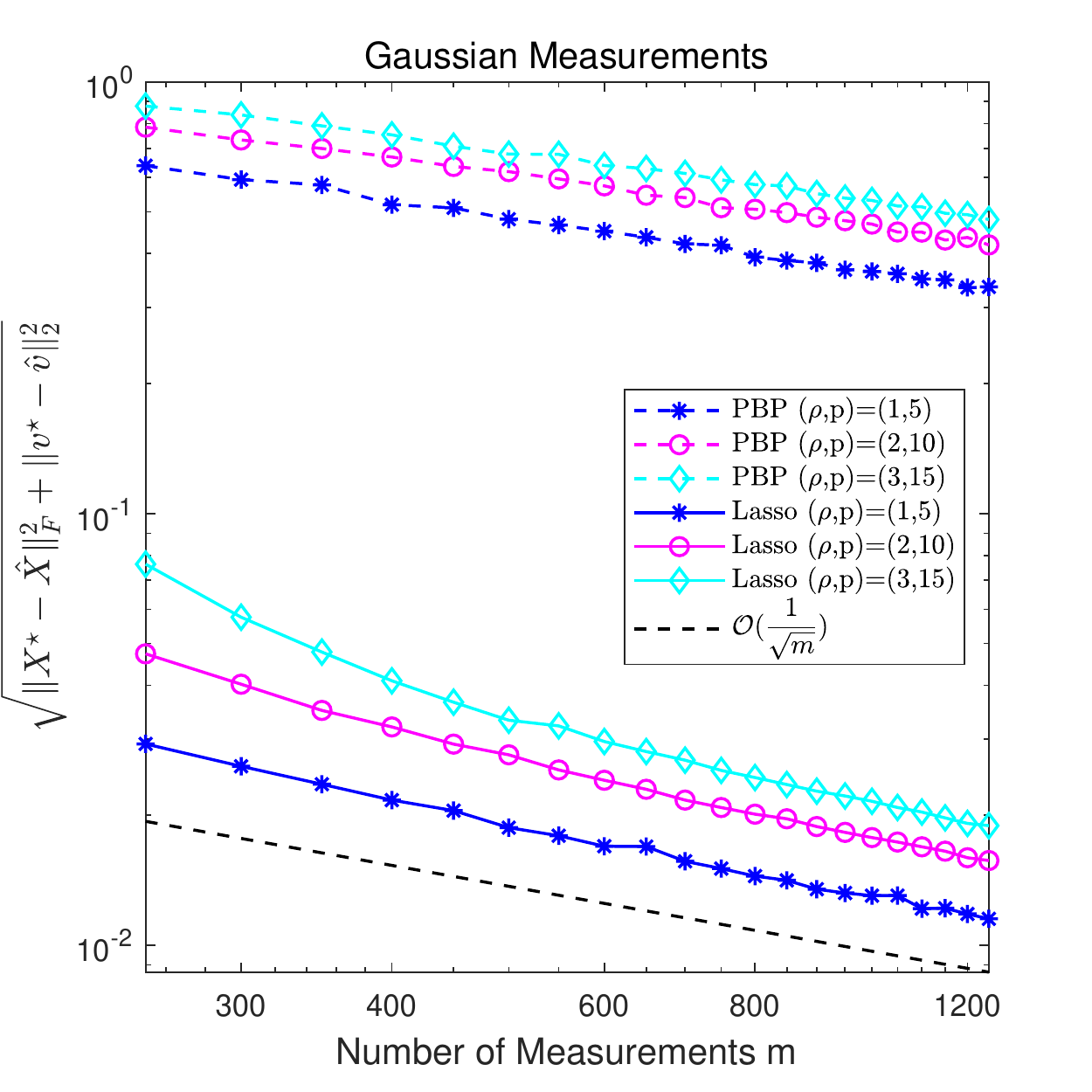}
		\label{fig3b}}
	\hfil
	\subfloat[Dependence on $\Delta$]{\includegraphics[width=2.0in]{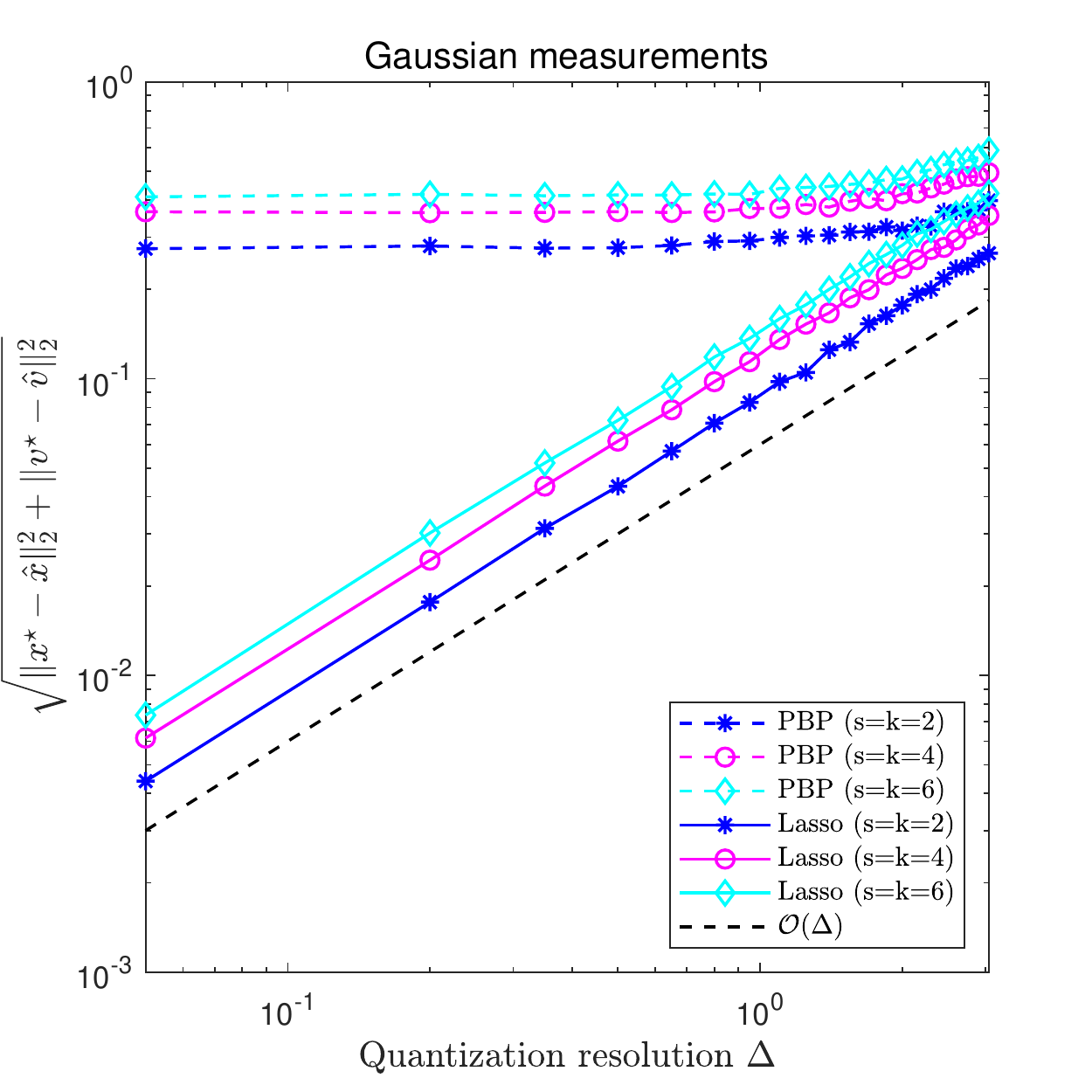}
		\label{fig3c}}
	\hfil
	\subfloat[Sparse signal recovery from sparse corruption]{\includegraphics[width=2.0in]{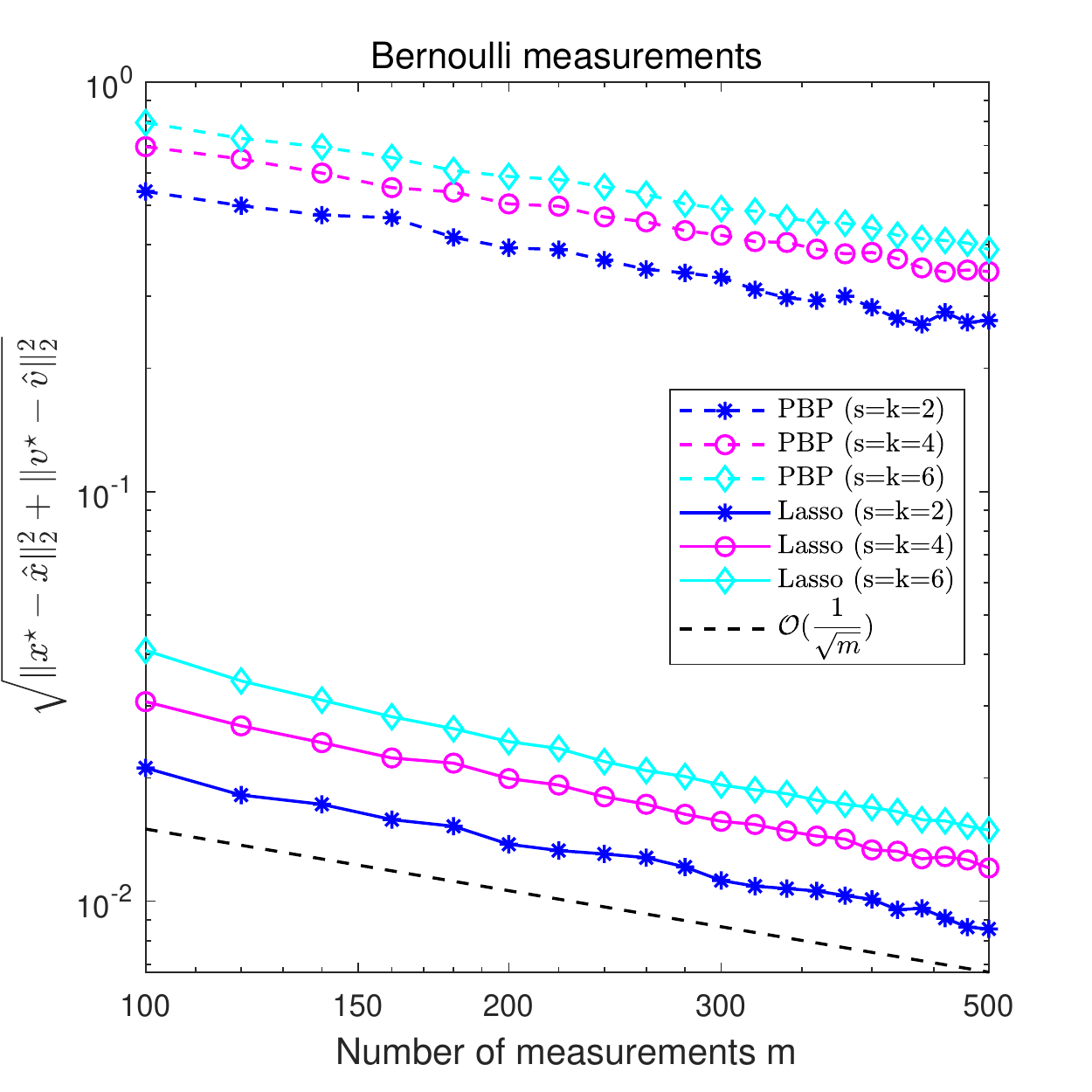}
		\label{fig3d}}
	\hfil
	\subfloat[Low-rank matrix recovery from sparse corruption]{\includegraphics[width=2.0in]{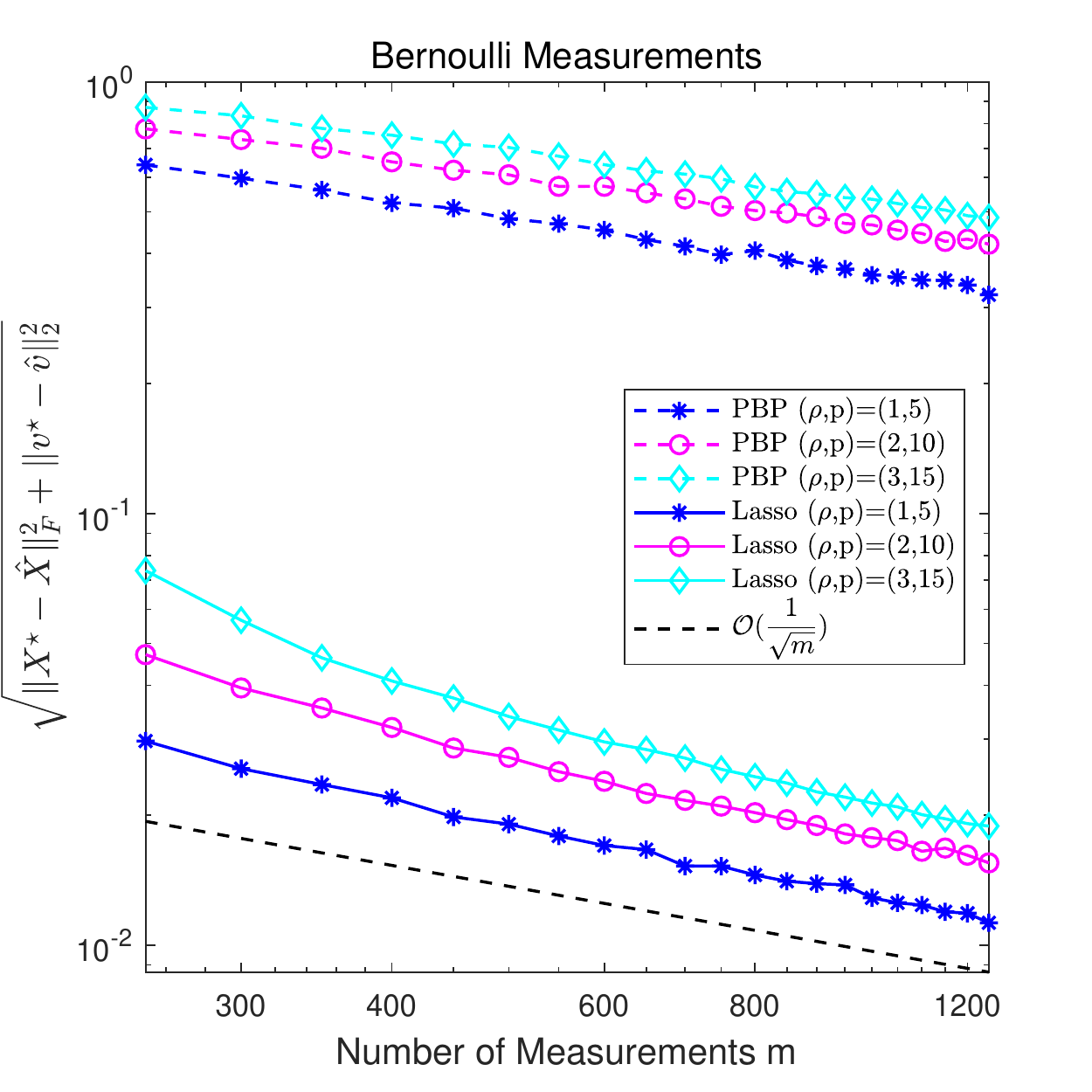}
		\label{fig3e}}
	\hfil
	\subfloat[Dependence on $\Delta$]{\includegraphics[width=2.0in]{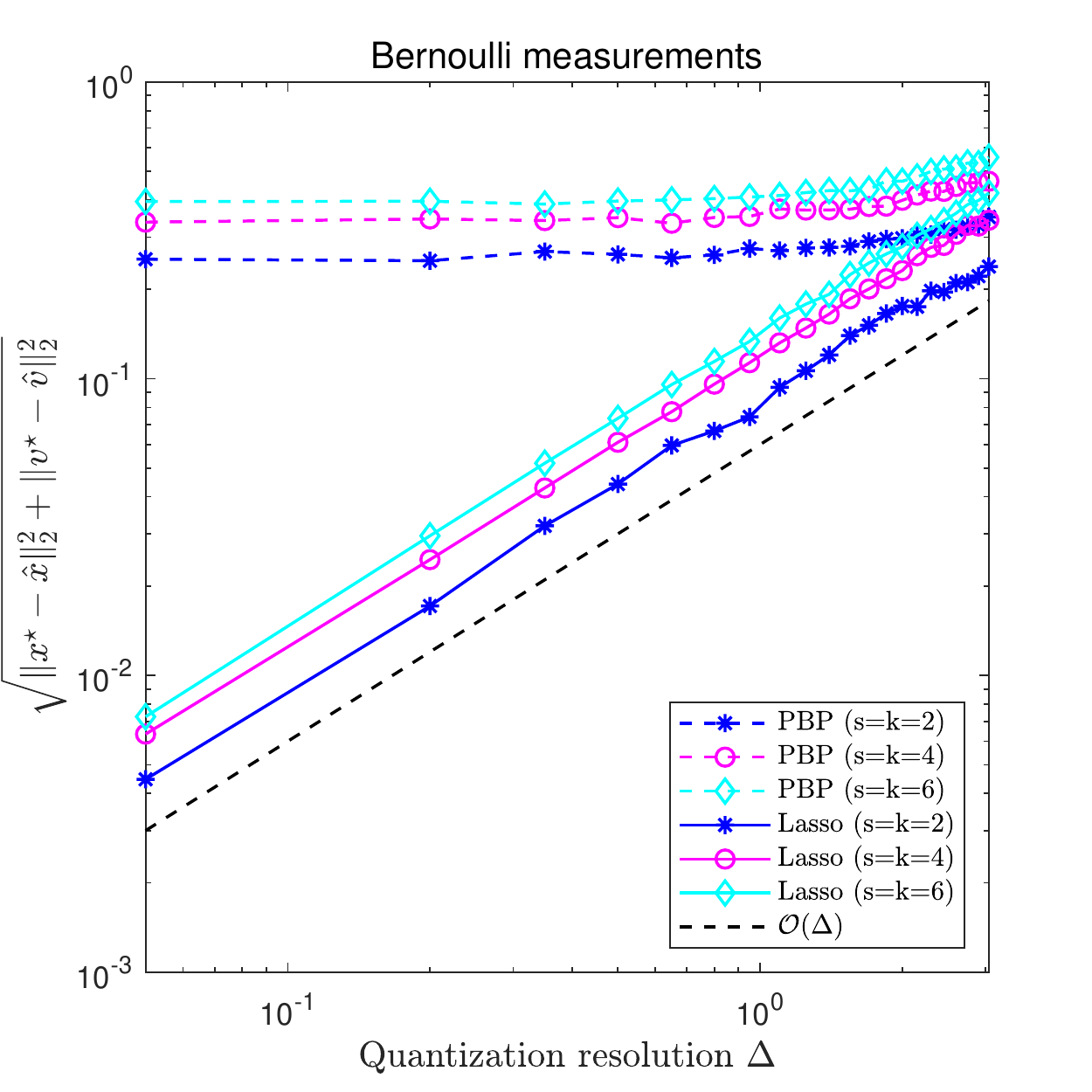}
		\label{fig3f}}
	\caption{Log-log error comparison between the constrained Lasso and the PBP method under Gaussian or Bernoulli measurements. The black dashed line illustrates the $\OO(\frac{1}{\sqrt m})$ and $\OO(\Delta)$ error scalings which are predicted in Theorem \ref{them: dither}.}
	\label{fig3}
\end{figure*}

\paragraph{Low-rank matrix recovery from sparse corruption}

Consider the case in which $\vx^{\star}=\text{vec}(\mX^{\star})\in\R^n$, where $\mX^{\star}\in\R^{d\times d}$ is a matrix with rank $\rho$ and $d^2=n$, and $\vv^{\star}$ is a $p$-sparse vector, then we obtain that the compatibility constants $\alpha_f=\sqrt{\rho}$ and $\alpha_g=\sqrt{p}$. Note that \cite[Exercise 10.4.3]{Vershynin2018}
\begin{align*}
\gamma(\B_f^n)=\gamma(\B_*^n)\leq 2 \sqrt{d},
\end{align*}
and
\begin{align*}
\gamma(\B_g^m)=\gamma(\B_1^m)\leq c\sqrt{\log m}.
\end{align*}
It then follows from \eqref{id of lbd1} and \eqref{id of lbd2} that we should choose
\begin{align*}
\lambda_1\geq C_2 K(\Delta+\epsilon)\sqrt{md},
\end{align*}
and
\begin{align*}
\lambda_2\geq C_2 K(\Delta+\epsilon)\sqrt{m\log m}.
\end{align*}

Similarly, to apply the bounds in Table \ref{table_bound2} conveniently, we pick $\kappa={2}/(C_2K(\Delta+\epsilon)\sqrt{m})$. Then we have ${\kappa}{\lambda_1}\geq2\sqrt{d}$ and ${\kappa}{\lambda_2}\geq\sqrt{2\log\frac{m}{p}}$, and hence
\begin{align*}
\eta^2\left(\kappa\lambda_1\cdot\partial \|\vx^\star\|_*\right)\leq {\kappa^2}{\lambda_1^2}\rho+2d(\rho+1),
\end{align*}
and
\begin{align*}
\eta^2\left(\kappa\lambda_2\cdot\partial \|\vv^\star\|_1\right)\leq \left({\kappa^2}{\lambda_2^2}+3\right)p.
\end{align*}

It follows from the upper bound \eqref{upper bound2} that
$$
\gamma^2(\CC_2\cap\S^{n+m-1}) \leq C'{\kappa^2} \cdot ({\lambda_1^2\rho+\lambda_2^2p})= C''\cdot\frac{\lambda_1^2\rho+\lambda_2^2p}{K^2(\Delta+\epsilon)^2m}.
$$
Combining the above bound with Theorem \ref{them: dither_FP} yields the following corollary.
\begin{corollary}\label{corollary_uncon_low-rank_recovery}
	Suppose that the signal $\mX^\star\in\R^{d\times d}$ is a $\rho$-rank matrix with $d^2=n$ and that the corruption $\vv^\star\in\R^m$ is a $p$-sparse vector. Let the regularization parameters satisfy $\lambda_1\geq C'K(\Delta+\epsilon)\sqrt{md}$ and $\lambda_2\geq C''K(\Delta+\epsilon)\sqrt{m\log m}$. Under the assumptions of Theorem \ref{them: dither_FP}, we have that, if the number of measurements
	$$
	m\geq c \cdot K\big({\lambda_1\sqrt{\rho}+\lambda_2\sqrt{p}}\big)/\big(\Delta+\epsilon\big),
	$$
	then, with high probability,
	\begin{align*}
	\sqrt{\|\hat{\mX} - \mX^{\star}\|_F^2+\|\hat{\vv} - \vv^{\star}\|_2^2}\leq C\cdot\frac{\lambda_1\sqrt{\rho}+\lambda_2\sqrt{p}}{m}.
	\end{align*}
\end{corollary}

\begin{remark}[Relationship between constrained and unconstrained Lassos]
	The above two examples have demonstrated that, by setting $\lambda_1,\lambda_2$ to their lower bounds in the unconstrained Lasso, one can approximately achieve the performance of the constrained Lasso. This should not be surprising since the theory of Lagrange multipliers \cite[Section 28]{rockafellar2015convex} asserts that the constrained and unconstrained procedures are essentially equivalent if one has chosen the regularization parameters $\lambda_1,\lambda_2$ correctly.
\end{remark}

\section{Numerical Simulations}\label{simulation}

In this section, we present a series of numerical experiments to verify our theoretical results. In all simulations, we draw the sensing matrices with standard normal entries for the Gaussian measurements, and symmetric Bernoulli entries for sub-Gaussian measurements. We solve the convex optimization problems by using the CVX Matlab package \cite{cvx,gb08}.

\subsection{Experiment Settings in Figure \ref{fig}}\label{fig_detail}
In this experiment, we intend to numerically show that direct application of generalized constrained Lasso \eqref{G-Lasso} to quantized corrupted  measurements \eqref{quantized corrupted sensing} (without dithering) cannot faithfully recover both signal and corruption. We consider $3$ kinds of nonlinearities: smooth hyperbolic tangent function, uniform quantization with resolution $\Delta=0.3$, and single bit measurement. The signal $\vx^{\star}\in\R^n$ is an $s$-sparse vector, the corruption $\vv^{\star}\in\R^m$ is a $k$-sparse vector or $\vv^{\star}=\vzero$. We normalize signal and corruption $\|\vx^{\star}\|_2=\|\vv^{\star}\|_2=1$ and generate the standard Gaussian $\mPhi$ to satisfy the unit-norm and Gaussian assumptions in \cite{Plan2015The}. We let $s=k=5$ and the signal dimension $n=128$. For each measurement, we carry out the constrained Lasso for 100 times. 

\subsection{Recovery Performance of Constrained Lasso}
\subsubsection{Sparse signal recovery from sparse corruption}
In this experiment, we consider the case in which both signal and corruption are sparse and the noise level $\epsilon=0$. The signal $\vx^{\star}\in\R^n$ is an $s$-sparse random vector whose supports are selected randomly, and nonzero entries are sampled i.i.d. from the standard Gaussian distribution. The corruption $\vv^{\star}\in\R^m$ is a $k$-sparse random vector which is drawn in the same way as $\vx^{\star}$. We fix the quantization resolution $\Delta=0.1$ and the signal dimension $n=256$, and vary the number of measurements $m$ between $100$ and $500$. For each pair of $(s,k)$, we run the experiment for 100 realizations. Figs. \ref{fig1a} and  \ref{fig1d} show the average error curves for each pair of $(s,k)$, the reconstruction error curves are consistent with the theoretical results under both Gaussian and Bernoulli measurements.

\subsubsection{Low-rank matrix recovery from sparse corruption}
We then investigate the recovery error when the signals are low-rank matrices. The noise level is set to $\epsilon=0$. Let $\mX^\star=\mU_1\mU_2^T$ be a $\rho$-rank matrix, where $\mU_1$ and $\mU_2$ are independent $d\times \rho$ matrices with orthonormal columns. The corruption signal $\vv^{\star}\in\R^m$ is a $p$-sparse random vector. We set the quantization resolution $\Delta=0.1$ and $d=16$, and vary number of measurements $m$ between $250$ and $1250$. We repeat the experiment 100 times for each pair of $(\rho,p)$. As shown in Figs. \ref{fig1b} and  \ref{fig1e}, the predicted $\OO(\frac{1}{\sqrt m})$ error scaling appears when the number of measurements $m$ exceeds a certain level.

\subsubsection{Robustness to noise}
In this experiment, we explore the empirical behavior of the recovery error under noisy measurements. We use almost the same experiment settings as the sparse signal recovery from sparse corruption case except that we add the truncated Gaussian noise. We first generate a standard Gaussian vector $\vz$ and then scale $\vz$ such that $\|\vz\|_{\infty}=\epsilon=[0.05, 0.1, 0.15]$. The sparsity level of signal and corruption is set to be $s=k=5$. Figs. \ref{fig1c} and   \ref{fig1f} plot the average error curves for 100 experiments. As shown in the figures, the recovery error is robust to the unstructured additive noise.

\subsection{Recovery Performance of Unconstrained Lasso}
Similarly, we carry out a series of experiments to investigate recovery error of the unconstrained Lasso. The simulation settings are nearly the same as the constrained case except that we use the unconstrained Lasso \eqref{f_penalized procedure} to reconstruct signal and corruption. The regularization parameters are set to their lower bounds in Corollaries \ref{corollary_uncon_sparse_recovery} and  \ref{corollary_uncon_low-rank_recovery}. As shown in Fig. \ref{fig2}, the recovery error curves behave as predicted by the theoretical results under both Gaussian and Bernoulli measurements.

\subsection{Performance Comparisons with the Other Approach}\label{performance_comparisons}
Finally, we compare the performance of the unconstrained Lasso \eqref{G-Lasso} with PBP \cite{Xu2018Quantized} for the recovery of signal and corruption from uniform dithered quantized measurements. In Figs. \ref{fig3a}, \ref{fig3b}, \ref{fig3d}, and \ref{fig3e}, we study the error dependence on the number of measurements $m$ under the similar experiment settings as in the previous subsections. For the PBP method, we solve
$(\hat{\vx},\hat{\vv})=\mP_\TT\left(\frac{1}{m}\bm{\Upsilon}^T\vy\right)$,
where $\bm{\Upsilon} = [\mPhi, \sqrt{m}\mI_m]$. In the sparse signal recovery from sparse corruption case, $\TT$ is set to $\TT=\{(\vx,\vv)\in\R^n\times\R^m: \|\vx\|_1\leq\|\vx^\star\|_1,\|\vv\|_1\leq\|\vv^\star\|_1\}$. In the low-rank matrix recovery from sparse corruption case, $\TT$ is set to $\TT=\{(\mX,\vv)\in\R^{d\times d}\times\R^m: \|\mX\|_*\leq\|\mX^\star\|_*,\|\vv\|_1\leq\|\vv^\star\|_1\}$. Figs. \ref{fig3a}, \ref{fig3b}, \ref{fig3d}, and \ref{fig3e} plot the recover error curves of the constrained Lasso and PBP under Gaussian and Bernoulli measurements. In both sparse signal and low-rank matrix recovery examples, the constrained Lasso significantly outperforms the PBP approach.

In Figs. \ref{fig3c} and \ref{fig3f}, we investigate the error dependence on quantization resolution $\Delta$. We consider the sparse signal recovery from sparse corruption case with the noise level $\epsilon=0$. We fix the number of measurements $m=500$ and vary the quantization resolution $\Delta$ between $0.05$ and $3.05$. As illustrated in Figs. \ref{fig3c} and \ref{fig3f}, the PBP observes an error floor at small values of $\Delta$, which is the error achieved by PBP in the unquantized measurements \cite{Xu2018Quantized}. On the contrary, the constrained Lasso enjoys a continuous error scaling $\OO(\Delta)$ as predicted in Theorem \ref{them: dither}.

\section{Conclusion}\label{Conclusion}
In this paper, we have investigated the problem of recovering a structured signal from quantized corrupted measurements. Under the dithered quantization framework, we have theoretically demonstrated that both constrained and  unconstrained Lassos can faithfully reconstruct signal and corruption with almost the same number of measurements as that in the linear case. We also illustrate how to choose regularization parameters for the unconstrained Lasso. Concrete examples and numerical simulations are provided to verify our theoretical results. For the future direction, it is interesting to explore the robust recovery of signal and corruption from other non-linear quantization schemes such as one-bit measurements and general non-linear sensing model.

\appendices
\section{Proofs of Main Results}\label{proof of main}

Before proving our main results, we require the following two useful lemmas.

\begin{lemma}[Quantization error of dithered quantizers] \cite[Theorem 1]{Gray1993Dithered}
	\label{quantization error}
	Let $\QQ_U(x)=\Delta(\lfloor\frac{x}{\Delta}\rfloor+\frac{1}{2})$ be the uniform quantizer. Suppose that $\{W_i\}_{i\in\mathbb{Z}}$ is a dithering signal which is independent of the input process $\{X_i\}_{i\in\mathbb{Z}}$ and is i.i.d.. The condition $\E\big(e^{juW_i}\big)|_{u=\frac{2\pi l}{\Delta}}=0$ for all $l\ne 0$ is necessary and sufficient for the following properties:
	\begin{itemize}
		\item $X_j$ is independent of the quantization error $\vz_i=\QQ_U(X_i+W_i)-(X_i+W_i)$ for all $i$ and $j$.
		\item The quantization error $\vz_i$ are i.i.d. uniform random variables on $(-\frac{\Delta}{2},\frac{\Delta}{2}]$.
	\end{itemize}
\end{lemma}

\begin{lemma}\label{lemma: UI}
	Let $\mPhi\in\R^{m \times n}$ be a matrix whose rows are independent centered isotropic sub-Gaussian vectors with $K = \max_i \|\mPhi_i\|_{\psi_2}$. Let $\vomega\in\R^m$ be a random bounded vector whose entries are mean-zero i.i.d. variables with $\|\vomega\|_{\infty}\leq \zeta$, and $\TT$ be a bounded subset of $\R^n\times\R^m$. Suppose that $\vomega$ is also independent of $\mPhi$. Then for any $t>0$, the event
	\begin{align*}
	\sup_{(\va,\vb)\in \TT} \ip{\mPhi\va + \sqrt{m}\vb}{\vomega} \leq CK\zeta\sqrt{m}\big[\gamma(\TT)+t\cdot\rad(\TT)\big]
	\end{align*}
	holds with probability at least $1-\exp(-t^2)$.
\end{lemma}

\begin{proof}
	See Appendix \ref{ProofofLemma2}.
\end{proof}

\subsection{Proof of Theorem \ref{them: dither}}

\begin{proof}
	For clarity, the proof is divided into three steps.
	
	\textbf{Step 1: Problem reduction.} Since $(\hat{\vx}, \hat{\vv})$ is the solution to the procedure \eqref{G-Lasso}, we have
	\begin{align}\label{reduction1}
	\|\vy-\mPhi\hat{\vx}-\sqrt{m}\hat{\vv}\|_2 \le \|\vy-\mPhi\vx^\star-\sqrt{m}\vv^\star\|_2.
	\end{align}
	Define the quantization error
	\begin{align*}
	\vz&:=\QQ_U(\bar{\vy}+\vtau)-(\bar{\vy}+\vtau) \\
	&=\QQ_U(\mPhi\vx^\star+\sqrt{m}\vv^{\star}+\vn+\vtau)-(\mPhi\vx^\star+\sqrt{m}\vv^{\star}+\vn)-\vtau.
	\end{align*}
	Let $\vh=\hat{\vx}-\vx^\star$ and $\ve=\hat{\vv}-\vv^\star$. Then \eqref{reduction1} can be reformulated as
	\begin{align}\label{reduction2}
	\|\mPhi\vh+\sqrt{m}\ve-\vz-\vn-\vtau\|_2 \le \|\vz+\vn+\vtau\|_2.
	\end{align}
	Squaring both sides of \eqref{reduction2} yields
	\begin{align}\label{reduction3}
	\left\| {\mPhi \vh + \sqrt m \ve} \right\|_2^2 \le 2\ip{\mPhi \vh + \sqrt m \ve}{\vz+\vn+\vtau}.
	\end{align}
	
	\textbf{Step 2: Lower Bound on $\| {\mPhi \vh + \sqrt m\ve}\|_2$.} Note that the error vector $(\vh,\ve)$ lies in the tangent cone $\CC_1$, which implies $\frac{1}{\sqrt{\|\vh\|_2^2+\|\ve\|_2^2}}\cdot(\vh,\ve)\in\CC_1\cap\S^{n+m-1}$. It then follows from Fact \ref{Ext MDI} (let $\TT=\CC_1\cap\S^{n+m-1}$ and choose $t=\gamma(\CC_1\cap\S^{n+m-1})$) that the event
	\begin{align*}
	&\|\mPhi \vh+ \sqrt m \ve\|_2\\
	&= \sqrt{\|\vh\|_2^2+\|\ve\|_2^2} \cdot \left\|\frac{\mPhi\vh}{\sqrt{\|\vh\|_2^2+\|\ve\|_2^2}}+\frac{\sqrt{m}\ve}{\sqrt{\|\vh\|_2^2+\|\ve\|_2^2}}\right\|_2  \\
	& \geq \sqrt{\|\vh\|_2^2+\|\ve\|_2^2} \cdot (\sqrt{m} - CK^2{\gamma(\CC_1\cap \S^{n+m-1})})\\
	& \geq \frac{{\sqrt m}}{2} \sqrt{\|\vh\|_2^2+\|\ve\|_2^2}
	\end{align*}
	holds with probability at least $1-\exp \{-\gamma^2(\CC_1\cap \S^{n+m-1})\}$. The last inequality is due to \eqref{NumberofMeasurements1}.
	
	\textbf{Step 3: Upper Bound on $\ip{\mPhi \vh + \sqrt m\ve}{\vz+\vn+\vtau}$.} Note that $\vtau_i\sim \text{Unif}(-\Delta/2,\Delta/2]$, and $$\E\big(e^{ju\vtau_i}\big)=\int_{-\Delta/2}^{\Delta/2}\frac{1}{\Delta}e^{jux}dx=\frac{\sin(u\Delta/2)}{u\Delta/2},$$
	hence $\E\big(e^{ju\vtau_i}\big)|_{u=\frac{2\pi l}{\Delta}}=0$ for all $l\ne 0$. By the definition of $\vz$ and Lemma \ref{quantization error}, $\vz_i=\QQ_U(\bar{\vy}_i+\vtau_i)-(\bar{\vy}_i+\vtau_i)$ are i.i.d. uniform variable on $(-\Delta/2,\Delta/2]$, and $\vz$ is also independent of $\bar{\vy}$ and hence is independent of $\mPhi$ and $\vn$. Then it follows Lemma \ref{lemma: UI} that (by setting $\vomega=\vz+\vn+\vtau,~\TT=\CC_1\cap \S^{n+m-1},~t=\gamma(\CC_1\cap \S^{n+m-1})$) the event
	\begin{align*}
	&\ip{\mPhi \vh + \sqrt m\ve}{\vz+\vn+\vtau}\\
	& = \sqrt{\|\vh\|_2^2+\|\ve\|_2^2} \ip{\frac{\mPhi\vh+\sqrt{m}\ve}{\sqrt{\|\vh\|_2^2+\|\ve\|_2^2}}}{\vz+\vn+\vtau}\\
	&\leq CK(\Delta+\epsilon)\sqrt{m} \sqrt{\|\vh\|_2^2+\|\ve\|_2^2} \cdot\gamma(\CC_1\cap \S^{n+m-1})
	\end{align*}
	holds with probability at least
	$1-\exp\{-\gamma^2(\CC_1\cap \S^{n+m-1})\}$. The last inequality holds because $\|\vomega\|_{\infty}\leq \Delta/2+\epsilon+\Delta/2$.
	
	Substituting the bounds of Steps 2 and 3 into \eqref{reduction3} and taking union bound, we have that, with probability at least
	$1-2\exp\{-\gamma^2(\CC_1\cap \S^{n+m-1})\}$,
	\begin{align*}
	&\frac{m}{4}(\|\vh\|_2^2+\|\ve\|_2^2)\\
	&\qquad \leq CK(\Delta+\epsilon)\sqrt{m} \sqrt{\|\vh\|_2^2+\|\ve\|_2^2} \cdot\gamma(\CC_1\cap \S^{n+m-1}).
	\end{align*}
	Rearranging completes the proof of Theorem \ref{them: dither}.
\end{proof}

\subsection{Proof of Theorem \ref{them: dither_FP}}

\begin{proof}
	The proof of Theorem \ref{them: dither_FP} is similar to that of Theorem \ref{them: dither} except for some modifications in the first step. Since $(\hat{\vx}, \hat{\vv})$ is the solution to the penalized procedure \eqref{f_penalized procedure}, we have
	\begin{align}\label{fp_reduction1}\notag
	\frac{1}{2}&\|\vy-\mPhi\hat{\vx}-\sqrt{m}\hat{\vv}\|_2^2+\lambda_1f(\hat{\vx})+\lambda_2g(\hat{\vv})\\
	&\le \frac{1}{2}\|\vy-\mPhi\vx^\star-\sqrt{m}\vv^\star\|_2^2+\lambda_1f(\vx^{\star})+\lambda_2g(\vx^{\star}).
	\end{align}
	Define the quantization error $\vz$ and error vectors $\vh,\ve$ as in the proof of Theorem \ref{them: dither}. Then \eqref{fp_reduction1} can be reformulated as
	\begin{align}\label{fp_reduction2}\notag
	&\frac{1}{2}\left\| {\mPhi \vh + \sqrt m \ve} \right\|_2^2 \le \ip{\mPhi \vh + \sqrt m \ve}{\vz+\vn+\vtau}\\
	&+\lambda_1\big(f(\vx^\star)-f(\vx^\star+\vh)\big)+\lambda_2\big(g(\vv^\star)-g(\vv^\star+\ve)\big).
	\end{align}
	Note that the left side of \eqref{fp_reduction2} is always nonnegative, then we have
	\begin{align*}
	&\lambda_1f(\vx^\star+\vh)+\lambda_2g(\vv^\star+\ve)\\
	&\qquad\leq \lambda_1f(\vx^\star)+\lambda_2g(\vv^\star)+\ip{\mPhi \vh + \sqrt m \ve}{\vz+\vn+\vtau}\\
	&\qquad\leq \lambda_1f(\vx^\star)+\lambda_2g(\vv^\star)+f^*\big(\mPhi^T(\vz+\vn+\vtau)\big)\cdot f(\vh)\\
	&\qquad\qquad+\sqrt{m}g^*(\vz+\vn+\vtau)\cdot g(\ve)\\
	&\qquad\leq\lambda_1f(\vx^\star)+\lambda_2g(\vv^\star)+\frac{\lambda_1}{2}\cdot f(\vh)+\frac{\lambda_2}{2}\cdot g(\ve),
	\end{align*}
	where $f^*(\cdot)$ and $g^*(\cdot)$ denote the dual norm of $f(\cdot)$ and $g(\cdot)$, respectively. The second inequality is due to H{\"o}lder's inequality. The last inequality follows from Condition \ref{condition2}. This further indicates that the error vector $(\vh,\ve)\in\CC_2$. Similar to Step $2$ in the proof of Theorem \ref{them: dither}, we have that the event
	\begin{align}\label{Lowbound}
	\|\mPhi \vh+ \sqrt m \ve\|_2\geq \frac{{\sqrt m}}{2} \sqrt{\|\vh\|_2^2+\|\ve\|_2^2}
	\end{align}
	holds with probability at least $1-\exp \{-\gamma^2(\CC_2\cap \S^{n+m-1})\}$.

   Note that the right side of \eqref{fp_reduction2} can be upper bounded
	\begin{align}\label{fp_upper_bound}\notag
	&\ip{\mPhi \vh + \sqrt m \ve}{\vz+\vn+\vtau}\\\notag
	&\qquad\quad+\lambda_1\big(f(\vx^\star)-f(\vx^\star+\vh)\big)+\lambda_2\big(g(\vv^\star)-g(\vv^\star+\ve)\big)\\\notag
	&\qquad\leq f^*\big(\mPhi^T(\vz+\vn+\vtau)\big)\cdot f(\vh)+\sqrt{m}g^*(\vz+\vn+\vtau)\\\notag
	&\qquad\qquad\cdot g(\ve)+\lambda_1\cdot f(\vh)+\lambda_2\cdot g(\ve)\\\notag
	&\qquad\leq\frac{\lambda_1}{2}\cdot f(\vh)+\frac{\lambda_2}{2}\cdot g(\ve)+\lambda_1\cdot f(\vh)+\lambda_2\cdot g(\ve)\\\notag
	&\qquad\leq\frac{3}{2}\cdot\left(\lambda_1\alpha_f\|\vh\|_2+\lambda_2\alpha_g\|\ve\|_2\right)\\
	&\qquad\leq\frac{3}{2}\cdot(\lambda_1\alpha_f+\lambda_2\alpha_g)\cdot\sqrt{\|\vh\|_2^2+\|\ve\|_2^2},
	\end{align}
	where $\alpha_f$ and $\alpha_g$ are compatibility constants. Here the first inequality is due to H{\"o}lder's inequality and the triangle inequality; the second inequality follows from Condition \ref{condition2}; the last inequality comes from the Cauchy-Schwarz inequality and the fact that $\sqrt{a^2+b^2}\leq a+b$ for all $a,b\geq 0$.

 Then, combining \eqref{fp_reduction2}, \eqref{Lowbound}, and \eqref{fp_upper_bound} yields the event
	\begin{align*}
	&\frac{m}{8}(\|\vh\|_2^2+\|\ve\|_2^2)\leq\frac{3}{2}\cdot(\lambda_1\alpha_f+\lambda_2\alpha_g)\cdot\sqrt{\|\vh\|_2^2+\|\ve\|_2^2}\\
	\end{align*}
   holds with probability at least $1-\exp\{-\gamma^2(\CC_2\cap \S^{n+m-1})\}$.

   Rearranging completes the proof of Theorem \ref{them: dither_FP}.
\end{proof}

\section{Proof of Lemma \ref{lemma: UI}}\label{ProofofLemma2}
Note first that $\vomega_i$ are bounded i.i.d. variables, by \cite[Exercise 2.5.8, Lemma 3.4.2]{Vershynin2018}, $\vomega$ is a sub-Gaussian random vector with    $$\|\vomega\|_{\psi_2} \leq  C_1 \zeta,~~\|\vomega\|_2 \leq \sqrt{m}\zeta.$$
	
	Define the random process $X_{\va,\vb}:=\ip{\mPhi\va+\sqrt{m}\vb}{\vomega}$, which has sub-Gaussian increments:
	\begin{align*}
	&\|X_{\va,\vb} - X_{\va',\vb'}\|_{\psi_2} = \left\|\ip{\mPhi(\va-\va')}{\vomega}+\sqrt{m}\ip{\vomega}{\vb-\vb'}\right\|_{\psi_2} \\
	&\qquad\leq \left\|\ip{\mPhi(\va-\va')}{\vomega}\right\|_{\psi_2}+\sqrt{m}\left\|\ip{\vomega}{\vb-\vb'}\right\|_{\psi_2} \\
	&\qquad\leq \|\vomega\|_2\cdot\|\mPhi(\va-\va')\|_{\psi_2} + \sqrt{m} \|\vb-\vb'\|_2 \cdot \|\vomega\|_{\psi_2} \\
	&\qquad\le C_2K\|\vomega\|_2\cdot\|\va-\va'\|_2 + C_1\zeta\sqrt{m}\cdot\|\vb-\vb'\|_2\\
	&\qquad\le C_2K\zeta\sqrt{m}\cdot\|\va-\va'\|_2 + C_1\zeta\sqrt{m}\cdot\|\vb-\vb'\|_2\\
	&\qquad\le C_3K\zeta\sqrt{m}\cdot\sqrt{\|\va-\va'\|_2^2+\|\vb-\vb'\|_2^2}
	\end{align*}
	for any $(\va,\vb),(\va',\vb')\in\TT$. The second inequality follows from the definition of sub-Gaussian norm of random vector and the third inequality is due to the fact that $\ip{\mPhi_i}{\va-\va'}$ are i.i.d. sub-Gaussian variables with $\|\ip{\mPhi_i}{\va-\va'}\|_{\psi_2}\leq K\|\va-\va'\|_2$ and hence $\|\mPhi(\va-\va')\|_{\psi_2} \leq CK\|\va-\va'\|_2$. The last inequality holds because $C_3K\ge C_1$ for some suitable absolute constant $C_3$.
	
	Define $\tilde{\TT}=\TT\cup\{(\vzero,\vzero)\}$, it then follows from Talagrand's Majorizing Measure Theorem (Fact \ref{Talagrand Them}) that the event
	\begin{align*}
	&\sup_{(\va,\vb)\in \TT} X_{\va,\vb} \leq \sup_{(\va,\vb)\in \TT} \left|X_{\va,\vb}\right| = \sup_{(\va,\vb)\in \tilde{\TT}} \left|X_{\va,\vb}\right|\\
	&=\sup_{(\va,\vb)\in \tilde{\TT}} \left|X_{\va,\vb}-X_{\vzero,\vzero}\right|
	\leq\sup_{(\va,\vb),(\va',\vb')\in \tilde{\TT}} \left|X_{\va,\vb}-X_{\va',\vb'}\right|\\
	&\leq C_3K\zeta\sqrt{m}\big[\omega(\tilde{\TT}) + t\cdot \diam(\tilde{\TT})\big]\\
	&\leq C_4K\zeta\sqrt{m}\big[\gamma(\TT) + t\cdot \rad(\TT)\big]
	\end{align*}
	holds with probability at least $1- \exp(-t^2)$. The last inequality holds because $\omega(\tilde{\TT})\leq\gamma(\tilde{\TT})=\gamma(\TT)$ and $\diam(\tilde{\TT}) = \sup_{\vx,\vy\in \tilde{\TT}}\|\vx-\vy\|_2\leq 2\rad(\tilde{\TT})=2\rad(\TT)$.

%
%

\bibliographystyle{IEEEtran}
\bibliography{IEEEabrv,myref}

\end{document}